\documentclass[
superscriptaddress,
nobibnotes,
 amsmath,amssymb,
 aps,
prb,
twocolumn,
]{revtex4-1}

\usepackage{graphicx}
\usepackage{dcolumn}
\usepackage{bm}
\usepackage{lipsum}
\usepackage{mathtools}
\usepackage{siunitx}
\usepackage{hyperref}
\usepackage{braket}
\hypersetup{
    colorlinks=true,
    linkcolor=blue,
    filecolor=magenta,      
    urlcolor=blue,
}

\newcommand{\hc}[1]{{#1}^{\dagger}}

\begin{document}
\title{Renormalized phonon spectrum of polyacetylene and similar materials}
\author{Stepan Fomichev}
\email{fomichev@physics.ubc.ca}
\affiliation{Department of Physics and Astronomy, University of
British Columbia, Vancouver BC V6T 1Z1, Canada} 
\affiliation{Stewart Blusson Quantum Matter Institute, University of
British Columbia, Vancouver BC V6T 1Z4, Canada}
\author{Mona Berciu}
\affiliation{Department of Physics and Astronomy, University of
British Columbia, Vancouver BC V6T 1Z1, Canada} 
\affiliation{Stewart Blusson Quantum Matter Institute, University of
British Columbia, Vancouver BC V6T 1Z4, Canada}
\date{\today}

\begin{abstract}
Motivated to understand the phonon spectrum renormalization in the ground state of the half-filled SSH model, we use the Born-Oppenheimer approximation together with the harmonic approximation to evaluate the all-to-all real-space ionic force constants generated through the electron-phonon interaction. Using these force constants, we compute the renormalized phonon spectrum and study its behaviour as a function of the Peierls distortion. For the undimerized chain we confirm the presence of a large Kohn anomaly at $2k_F$, signalling a strong lattice instability. For the dimerized chain, we find an optical branch separated by a gap from the acoustic one, while the Kohn anomaly manifests as phonon softening. To find the equilibrium dimerization, we minimize the ground state energy, crucially including the contribution of the renormalized phonon zero-point energy (ZPE). Our results show strong agreement with prior \textit{ab initio} studies for trans-polyacetylene and linear acetylenic carbon (carbyne), validating our method which is much simpler, and moreover can be easily generalized to study other problems in higher dimensions. 
\end{abstract}

\maketitle

\section{Introduction}
\label{sec:intro}

The concept of Peierls instability in one-dimensional (1D) metals -- the idea that 1D lattices are prone to dimerization in the presence of electron-phonon coupling -- has been a source of inspiration (and perspiration) for the condensed matter community for over half a century. The extent to which it actually occurs in real materials such as (trans-)polyacetylene is particularly interesting, especially as organic conductors grow in technological relevance (\textit{e.g.} in the fields of organic photovoltaics, field-effect transistors). Indeed, understanding the origin of conductivity (or semiconductor behaviour) in plastics can enable great improvements in their characteristics due to their extreme tunability. 

Early pre-war calculations of Lennard-Jones \cite{Lennard-Jones1937} and Coulson \cite{Coulson1939} using LCAO molecular orbital theory predicted that as $n \rightarrow \infty$ in finite-length polyene chains C$_{2n}$ H$_{2n+2}$ (\textit{i.e.} approaching the polyacetylene limit), the differences in bond length (and thus any electronic gap) would go to zero -- in line with the experimental observations available at the time \cite{Kuhn1937}. However, already in the 1950s it was becoming apparent that bond length alternation does persist even in the infinite length limit in 1D chains, with the arguments of Peierls \cite{Peierls2001}, Ooshika \cite{Ooshika1957} and Longuiet-Higgins and Salem \cite{Longuet-Higgins1959} and, finally, of Su, Schrieffer and Heeger (SSH) \cite{Su1979} declaring trans-polyacetylene to be a semiconductor. 

However, these approaches hinged on mean-field arguments and largely neglected the quantum nature of the crystal lattice. In the 1980s it became clear that the magnitude of the zero-point fluctuations $\Delta u$ in these 1D chains is comparable to the size of the dimerization \cite{Su1982}. Moreover, the so-called Peierls barrier -- the energy difference between the dimerized and undimerized structure (calculated based on extrapolations from finite length polyenes) -- is smaller than the zero-point fluctuation energy. These observations called into question the validity of neglecting the quantum nature of the lattice. 

To address these concerns, Nakahara and Maki \cite{Nakahara1982} evaluated corrections to the SSH result using the continuum version of the SSH model and found that dimerization indeed survives this challenge, albeit with a reduced dimerization value (Su found the same result using Monte Carlo simulations in Ref. \onlinecite{Su1982}). To see if this conclusion also applies in the non-adiabatic regime, Fradkin and Hirsh  carried out a fully quantum-mechanical calculation (on a continuum model) using a numerical renormalization group (NRG) technique, and found that even with a quantized lattice, dimerization will occur for any finite value of electron-phonon coupling, zero-point fluctuations notwithstanding\cite{Fradkin1983}.

In the 2000s and 2010s, researchers became interested in Peierls dimerization on a lattice, including electron-electron interactions with the Hubbard model. Various approaches using NRG, continuous-time quantum Monte Carlo, density matrix renormalization group (DMRG) and other numerical methods were successfully employed by Sengupta \text{et al.} \cite{Sengupta2003}, Barford and Pearson \textit{et al.} \cite{Barford2006, Pearson2011}, Weber \textit{et al.} \cite{Weber2015} and others: they demonstrated that the half-filled spinful Hubbard-SSH model with quantum acoustic phonons (the best approximation to real polyacetylene) admits two ground states, a Mott-Hubbard insulator and a Peierls (dimerized) insulator, but no metallic phase. Thus materials like polyacetylene can be expected to exhibit robust dimerization for arbitrary strength of the electron-phonon coupling even in the presence of relatively large zero-point fluctuations. Furthermore, their best estimates of the parameters appropriate for polyacetylene within these models (such as those estimated from a Pariser-Parr-Pople-Peierls model \cite{Barford2002}) place polyacetylene within the Peierls phase, although fairly close to the phase line separating the two. Recent experimental measurements for very long linear acetylenic carbon chains with $N \sim 6000$ (carbyne, the $-$C$\equiv$C$-$ carbon allotrope) appear to confirm the dimerization hypothesis \cite{Shi2016}.

And yet in recent years a number of works have questioned whether quantum lattice effects are indeed as small as the model Hamiltonian studies would suggest. In their 2013 publication, Hudson \textit{et al.}\cite{Hudson2013} argue against the bond length alternation hypothesis for polyacetylene and instead claim that infinite-length polyacetylene is metallic, invoking criticism of existing experiments on polyacetylene and supporting \textit{ab initio} DFT calculations with the PBE0 functional and a single-particle Fourier Grid Hamiltonian (FGH) method for the ionic Hamiltonian. Their \textit{ab initio} findings appear to be supported by other researchers \cite{Artyukhov2014}, as well as some recent experimental transport studies on short ($N = 10$ to $20$) carbyne chains \cite{LaTorre2015,BenRomdhane2017}, which also found that un-strained carbon chains are metallic. 

According to Hudson \textit{et al.}, the bond length alternation hypothesis for polyacetylene rests on the reasoning that ``since polyacetylene is not a conductor, it must be a semiconductor, and thus it must exhibit bond alternation'' \cite{Hudson2018}. Polyacetylene's insulating and Raman spectra properties could be explained with electron-electron interactions without recourse to lattice dimerization and Peierls instability, as already suggested by Ovchinnikov and co-workers in the early 1980s \cite{Ovchinnikov1982}. Moreover, in their two-step calculation with DFT and single-particle FGH, they find that the zero-point motion is sufficiently strong to favor undimerized chains.

Hudson \textit{et al.} also advanced another potential cause for the poor conductivity of commonly used polyacetylene: most samples of polyacetylene, traditionally synthesized at an elevated temperature which favours trans- (as opposed to cis-) polyacetylene formation, are not the infinite one-dimensional chains modeled in theoretical studies. Rather, all such samples involve a ``mixture of finite chains and cross-linked polymers'' \cite{Hudson2018}, which result in poor conductivity due to ``end effects''. In fact, many of the other properties of polyacetylene that are measured in experiments such as X-ray diffraction (space group analysis) \cite{Zhu1992}, nuclear magnetic resonance spectroscopy (single/double bond lengths, $sp^2$/$sp^3$ bond character) \cite{Yannoni1983}, Resonance Raman spectroscopy (vibrational frequencies) \cite{Schen1988}, have an alternative interpretation as consequences of polyacetylene samples being in reality mixtures of polyenes, or short-length chains of the repeating C$_2$H$_2$ unit. Hudson and others are currently working on a different way of synthesizing polyacetylene, relying on urea inclusion complexes to produce long, non-cross-linked, quasi-1D chains \cite{Dinca2020}.

Fundamentally, then, it appears that there are still a number of questions to be answered about the phonon spectrum and the importance of quantum mechanical effects of the lattice to the nature of the ground state. 

There are several approaches to quantifying the effect of the electron-phonon interaction on the phonon spectrum. Early work by Ovchinnikov \textit{et al.}\cite{Ovchinnikov1973,Ovchinnikov1982} has focused on mean-field semiclassical approaches, and found that electron-electron interaction inclusion was crucial for correctly reproducing the observed Raman excitation frequencies for polyacetylene. They were also highly skeptical of the dimerization hypothesis, suggesting that there is nothing in the data that cannot be explained by including the electron interactions, and in fact including dimerization would significantly overestimate the softening of the optical phonon.

Others, such as Nakahara and Maki \cite{Nakahara1982} and Schulz \cite{Schulz1978}, used Green's function methods and the random phase approximation (RPA) to calculate the phonon spectrum renormalization, demonstrating phonon softening and phonon gap opening at the Brillouin zone edge. However, they never explicitly sought to calculate the ground state dimerization by minimizing the total system energy (including the zero-point energy), as we do in this study -- the lattice structure was always prescribed ahead of time and never treated as an adjustable parameter.

Most recently, there has been an explosion of high-fidelity numerical methods, based on DFT, \textit{ab initio} molecular dynamics (AIMD), as well as advanced variational methods such as self-consistent \textit{ab initio} lattice dynamics (SCAILD)\cite{Souvatzis2008} and stochastic self-consistent harmonic approximation (SSCHA)\cite{Monacelli2021} (see the introduction in Ref. \onlinecite{Monacelli2021} for a comprehensive recent review). The incredible accuracy of such heavy numerical approaches can sometimes come at the expense of physical insight that can be more easily extracted from simple Hamiltonian model-based approaches.

In this paper we propose a straightforward approach for calculating the effects of electronic behaviour on the phonon spectrum of a 1D chain, using a combination of the Born-Oppenheimer approximation, the harmonic approximation, and perturbation theory. We apply this formalism to the standard SSH Hamiltonian in the non-adiabatic limit. 

Using this technique, we confirm that the phonon spectrum of the undimerized chain acquires a large Kohn anomaly at twice the Fermi wavevector $q_c = 2k_F$, indicating that the lattice is indeed unstable to dimerization, even with the explicit inclusion of zero-point motion effects. Through the interaction with the extended electron states, longer range force constants arise between ions, even when the bare forces are nearest-neighbour only. Therefore, even though we start with just an acoustic phonon branch, once we calculate the impact of the coupling to the electrons and minimize the total system energy, the phonon spectrum evolves an independent optical branch, separated from the acoustic one by a gap at the Brillouin zone edge. The optical branch retains some $\Gamma$-point softening from the Kohn anomaly even in the dimerized ground state. Our results allow us to address some of the confusion around the Peierls instability and bond length alternation in polyacetylene and related carbon chains. We are also able to shed light on the discrepancy between the two opposing predictions regarding the presence/absence of dimerization of the lattice, and to re-interpret the zero-dimerization experimental results.

The work is organized as follows: in Section \ref{sec:model}, we describe our
effective model Hamiltonian. Section \ref{sec:BO-calc} reviews, in broad strokes, the techniques used to study it (details are delegated to various appendices). Our results are discussed in Section \ref{sec:res-and-disc}, and we conclude with some final remarks in Section \ref{sec:concl}.

\section{The Model}\label{sec:model}

We start from a Hamiltonian describing $\pi$ electrons at half-filling (one electron per site) in a one-dimensional chain with lattice constant $a$
\begin{multline}\label{eq:ham}
H = -\sum_{n\sigma} t_{n,n+1} \left( \hc{c}_{n+1,\sigma} c_{n\sigma} + \text{h.c.}  \right) + \\
+ \sum_n \frac{\hat{p}_n^2}{2M} + \frac{K}{2} \sum_n (\hat{u}_{n+1} - \hat{u}_n)^2.
\end{multline}

Here $\hc{c}_{n\sigma}$ is a creation operator for an electron on site $n$ with spin $\sigma$, $t_{n,n+1}$ are nearest-neighbor hopping integrals, $K$ is the stiffness of the $\sigma$ bonds between neighboring ions, and $\hat{u}_n = \hat{R}_n - na$ are the (operator) deviations of the ions from the undistorted equilibrium positions. 

This Hamiltonian neglects electron-electron interactions, in line with some previous investigations \cite{Su1979}. Other standard approximations are to include only nearest-neighbor hopping and only nearest-neighbor effective interactions between ions, plus the reduction of the geometric complexity of real chain polymers (the zig-zag structure of trans-polyacetylene, presence of H or other ligands, out-of-plane bending and torsion) down to a one-dimensional chain. These approximations are justified to one degree or another\cite{Ovchinnikov1982}, but can also be relaxed and treated within the approach we propose below. In particular, electron-electron interactions (which may be important, see eg. Ref. \onlinecite{Ovchinnikov1973}) can be straightforwardly added within a Hartree-Fock treatment.

Next, we adopt the SSH prescription for obtaining the electron-phonon coupling from $t_{n,n+1} = t - \alpha (\hat{u}_{n+1} - \hat{u}_n)$ (more discussion is in Appendix \ref{app:quad-SSH-details}), which leads to the well-known SSH model:
\begin{multline}\label{eq:full-problem}
H = -t \sum_{n\sigma} \left( \hc{c}_{n+1,\sigma} c_{n\sigma} + \text{h.c.}  \right) + \\
+ \alpha \sum_{n\sigma} (\hat{u}_{n+1} - \hat{u}_n) \left( \hc{c}_{n+1,\sigma} c_{n\sigma} +  \text{h.c.}  \right) + \\
+ \sum_n \frac{\hat{p}_n^2}{2M} + \frac{K}{2} \sum_n (\hat{u}_{n+1} - \hat{u}_n)^2.
\end{multline}

Most studies of polyacetylene using this SSH model are in the adiabatic limit $M \rightarrow \infty$, and the parameter values ($t, \alpha, K$) are those from the original SSH paper \cite{Su1979}, up to minor variations. Those values were obtained by adjusting the parameters to reproduce the contemporary measurements of electronic bandgap and bond length alternation. Given that electron-electron interactions were neglected and that further experimental measurements have since become available (see Refs. \onlinecite{Hudson2018, Swager2017} for reviews), the true set of model parameters representing real-world polyacetylene should perhaps be revisited. This becomes even more important if, as discussed in the Introduction, polyacetylene samples are contaminated with a distribution of short-length polyenes, which could affect bandgap and bond length alternation.  Given the difficulties in determining the appropriate parameter values, we instead investigate a whole range of possibilities for ($t, \alpha, K,M$): we only demand that $\alpha ||\hat{u}_n|| \ll t$ and $||\hat{u}_n|| \ll a$, so that the harmonic approximation remains valid.

\section{Calculation}\label{sec:BO-calc}

\subsection{Born Oppenheimer decomposition}

In this section, we present our treatment of the SSH Hamiltonian within the Born-Oppenheimer approximation, which is justified given the small but finite ratio of the electron and ion masses.  This separation between the  mass scales allows us to solve separate electronic and ionic problems sequentially, with the total electronic energy acting as a potential for the ions (that is, defining a Born-Oppenheimer potential energy surface). For clarity, we now sketch the Born-Oppenheimer procedure. 

Consider a generic Hamiltonian 
\begin{equation}
\hat{H} = \underbrace{\hat{T}_e  + \hat{V}_{e-i} + \hat{V}_{i-i}}_{ = \hat{H}_{e}} + \hat{T}_i + \underbrace{\hat{V}_{e-e}}_{\text{neglect}},
\end{equation}
where $\hat{T}_{e/i}$ are the kinetic energies of electrons ($e$) and ions ($i$), and $\hat{V}$ are the various Coulomb interactions, respectively. (As discussed, in this work we will neglect the electron-electron interactions, however the Born-Oppenheimer analysis holds in their presence, as well.) We aim to solve the Schr\"odinger equation
\begin{equation}
\hat{H} \Psi(\{R_n, r_n\}) = E \Psi(\{R_n, r_n\})
\end{equation}
where $R_n$ are the ions' positions, and $r_n$ are the electrons' positions (for simplicity, we restrict these to one dimension). Following Born and Oppenheimer, we assume that the  wavefunction can be factorized: $\Psi(R_n, r_n) \equiv \psi(r_n; R_n) \phi(R_n)$ (note that a more complicated ansatz of Born-Huang type \cite{Born1954} is possible). The electronic component $\psi$ satisfies the Sch\"odinger equation
\begin{equation}\label{eq:schro-full}
\hat{H}_{e} \psi(r_n; R_n) = E_{e}(R_n) \psi(r_n; R_n)
\end{equation} 
and depends on the ion positions' $R_n$ as parameters.
The ionic component $\phi$ satisfies the equation
\begin{equation}\label{eq:BO-phonon}
\Big[ \hat{T}_i + E_{e}(\hat{R}_n) \Big] \phi(R_n) = E \phi(R_n).
\end{equation}

Given these assumptions, we have:
\begin{multline}\label{eq:BO-key}
\hat{H} \psi(r_n; R_n) \phi(R_n) 
= \psi(r_n; R_n) \Big[ E_{e}(R_n) + \hat{T}_i \Big] \phi(R_n) + \\
+ \Big[ \hat{T}_i \psi(r_n ; R_n) \Big] \phi(R_n)  \\
= E \psi(r_n; R_n) \psi(R_n) +  \Big[ \hat{T}_i \psi(r_n ; R_n) \Big] \phi(R_n). 
\end{multline}
The last term in  Eq. \ref{eq:BO-key} is the term preventing the Born-Oppenheimer ansatz from being exact. An argument due to Slater\cite{Slater1951} demonstrates this term to be smaller than the rest by a factor of $m_e/M$: note that the last term is $\hbar^2/(2M) \partial^2\psi/\partial R_n^2 \approx \hbar^2/(2M) \partial^2\psi/\partial r_n^2$. The approximate equality holds because the wavefunction depends on the differences $r_n - R_n$ of the coordinates and thus both derivatives remain of the same order. However, $E_e$ will have in it the kinetic energy of the electrons, which is order $\hbar^2/(2m_e) \partial^2\psi/\partial r_n^2$, with the difference being precisely a factor of $m_e / M$. Therefore ignoring the last term in Eq. \ref{eq:BO-key} is an accurate approximation for $m_e \ll M$.

Applying this approximation to the SSH Hamiltonian in Eq. \ref{eq:full-problem}, it factorizes into effective electronic and ionic components as defined by:
\begin{multline}\label{eq:BO-elec}
H_{\text{e}} = -t \sum_{n\sigma} \left( \hc{c}_{n+1,\sigma} c_{n\sigma} + \text{h.c.}  \right) + \\
+ \alpha \sum_{n\sigma} \langle \hat{u}_{n+1} - \hat{u}_n \rangle \left( \hc{c}_{n+1,\sigma} c_{n\sigma} +  \text{h.c.}  \right) +\\
+  \frac{K}{2} \sum_n \langle \hat{u}_{n+1} - \hat{u}_n \rangle^2,
\end{multline}
\begin{equation}\label{eq:BO-ion}
H_{i}^{(\text{BO})} = \sum_n \frac{\hat{p}_n^2}{2M} + E_e(\{\hat{R}_n\}).
\end{equation}
where $\langle \hat{u}_n\rangle \equiv R_n- na $.

\subsection{Electronic energy}

Keeping in line with previous work going back to Peierls \cite{Peierls2001}, we expect a dimerization of the one-dimensional chain. To single out this dominant static distortion from other, likely smaller, lattice fluctuations, we adopt the ansatz
\begin{equation}
\hat{u}_n = (-1)^n u + \hat{x}_n, \quad  \hat{x}_n  \ll  u.
\end{equation}
We will check the self-consistency of this approximation {\it a posteriori}, as well as minimize the total energy with respect to this variational parameter which sets the new equilibrium positions to $R_n^0=na +(-1)^n u$.

Combining the static distortion term with the hopping term, we obtain the standard electronic SSH solution plus a perturbation (we define  $\langle x_n \rangle = x_n$)
\begin{multline}
H_e = - \sum_{k\sigma} E_k(u) ( \hc{\nu}_{k\sigma} \nu_{k\sigma} - \hc{\chi}_{k\sigma} \chi_{k\sigma} ) + \\
+ \alpha \sum_{n\sigma} (x_{n+1} - x_n ) \left( \hc{c}_{n+1,\sigma} c_{n\sigma} +  \text{h.c.}  \right) + \\
+ \frac{K}{2} \sum_n (x_{n+1} - x_n - 2 (-1)^n u)^2.
\end{multline}

The first term involves the usual conduction and valence band operators:
\begin{align}
\nu_{k\sigma} &= \alpha_k c_{k\sigma}^{(e)} + \beta_k c_{k\sigma}^{(o)} \\
\chi_{k\sigma} &= \alpha_k c_{k\sigma}^{(e)} -  \beta_k c_{k\sigma}^{(o)}
\end{align}
where $\alpha_k = \frac{1}{\sqrt{2}}, \beta_k = \frac{1}{\sqrt{2}} \left(\frac{\epsilon_k + i u \Delta_k}{E_k(u)} \right)$,
\begin{align}
\epsilon_k &= -2t\cos(ka)\\
\Delta_k &= 4\alpha \sin(ka),
\end{align}
and
\begin{align}
    {c^{(o)}_{k\sigma}} &= \sum_n \frac{e^{-i k(2n+1)a}}{\sqrt{N}} {c_{2n+1,\sigma}} \\
{c^{(e)}_{k\sigma}} &= \sum_n \frac{e^{-i k(2n)a}}{\sqrt{N}} {c_{2n,\sigma}}.
\end{align}
Furthermore, the band energies are defined by 
\begin{equation}
E_k(u) = \sqrt{\epsilon_k^2 + u^2\Delta_k^2}.
\end{equation}
This first term is the usual SSH result, showing a gapped electronic spectrum whose size is controlled by the static dimerization $u$. 

The equilibrium contributions ($x_n = 0$) to the ionic Hamiltonian, at half-filling, are:
\begin{multline}
\label{eq:SSH_energy}
E_{e}(\{R_n^0\}) = -2 \sum_k E_k(u) + 2Ku^2 N = \\
= - \frac{Na}{\pi} \int_{-\frac{\pi}{2a}}^{\frac{\pi}{2a}} \, dk \sqrt{ \epsilon_k^2 + u^2\Delta_k^2 } + 2 K u ^2 N.
\end{multline}

The next step is to find the dependence of the electronic energy on all $x_n= R_n - R_n^0$, so that we can then use $E_e(\{\hat{R}_n\})$ to solve the ionic problem. 

\subsection{Harmonic approximation}

To allow further analytical progress, we use the harmonic approximation to deal with the Born-Oppenheimer potential energy surface $E_e(\{\hat{R}_n\})$. The zeroth order term is just the (electronic) energy associated with the equilibrium lattice positions, $E_e(\{R_n^0\})$, listed above. The first order term disappears as we are expanding the electronic energy around the new  equilibrium lattice positions. To second order in ionic displacements,  we are then left with: 
\begin{equation}\label{eq:BO-basic}
H_{i,\text{har}}^{(\text{BO})} = \sum_n \frac{\hat{p}_n^2}{2M} + E_e(\{ R_n^0 \}) + \frac{1}{2} \sum_{nm} \frac{\partial^2 E_e}{\partial x_n \partial x_m} \hat{x}_n \hat{x}_m.
\end{equation}
The second-order derivatives are commonly known as the dynamical matrix. We now proceed to calculate them.

Substituting Eq. (\ref{eq:SSH_energy}) into Eq. (\ref{eq:BO-basic}), we find
\begin{multline}\label{eq:BO-basic2}
H_{i,\text{har}}^{(\text{BO})} = \sum_n \frac{\hat{p}_n^2}{2M} + 2 K u^2 N - 2 \sum_{k} E_k + \\
+\frac{1}{2} \sum_{nm} \frac{\partial^2 E_e}{\partial x_n \partial x_m} \hat{x}_n \hat{x}_m.
\end{multline}

An immediate simplification can be made at this stage, by passing the $\sigma$-bond quadratic terms through the second-order derivatives of the electronic energy (they do not depend on the electronic wavefunctions and thus the expectation value over them gives the identity)
\begin{multline}\label{eq:BO-basic3}
H_{i,\text{har}}^{(\text{BO})} = \sum_n \frac{\hat{p}_n^2}{2M} + \frac{K}{2} \sum_n (\hat{x}_{n+1} - \hat{x}_n)^2 + 2 K u^2 N - \\
- 2 \sum_{k} E_k + \frac{1}{2} \sum_{nm} \frac{\partial^2 \langle \hat{H}_e - \hat{V}_{i-i} \rangle}{\partial x_n \partial x_m} \hat{x}_n \hat{x}_m.
\end{multline}
where $\hat{V}_{i-i} = (K/2) \sum_n (\hat{x}_{n+1} - \hat{x}_n - 2 (-1)^n u)^2$. 

To evaluate these derivatives, we use perturbation theory to calculate the expectation value $F_e \equiv \langle \hat{H}_e - \hat{V}_{i-i} \rangle$ to second order in $x_n$. All the non-trivial dependence comes from the electron-phonon coupling term, which we label $\hat{U}_{el-ph}$.
To set the stage for the perturbative calculation, we rewrite $\hat{U}_{el-ph}= \sum_n x_n \hat{f_n}$ where the electronic operators $\hat{f}_n$  are expressed in terms of the $\nu_{k,\sigma}$ and $\chi_{k,\sigma}$ conductance and valence band operators. Their form depends on whether $n$ is even or odd. Specifically, using the shorthand $\sin(ka) \equiv s_k$, we find
\begin{widetext}
\begin{align}\label{eq:V-even}
\hat{f}_{2n} =& \frac{ \alpha}{N/2} \sum_{kq\sigma} (2i) e^{-i(k-q)(2n)a}\begin{pmatrix}
\hc{\nu}_k & \hc{\chi}_k
\end{pmatrix}
\begin{pmatrix}
-\alpha_k (s_k \beta_k - s_q \beta_q^*) & - \alpha_k (s_k \beta_k + s_q \beta_q^*) \\
\alpha_k ( s_k \beta_k + s_q \beta_q^*) & \alpha_k (s_k \beta_k - s_q \beta_q^*)
\end{pmatrix}
\begin{pmatrix}
\nu_q \\
\chi_q
\end{pmatrix}, \\
\label{eq:V-odd}
\hat{f}_{2n+1} =& \frac{ \alpha}{N/2} \sum_{kq\sigma} (2i) e^{-i(k-q)(2n+1)a} \begin{pmatrix}
\hc{\nu}_k & \hc{\chi}_k
\end{pmatrix}
\begin{pmatrix}
\alpha_k (s_q \beta_k - s_k \beta_q^*) & \alpha_k (s_q \beta_k + s_k \beta_q^*) \\
-\alpha_k ( s_q \beta_k + s_k \beta_q^*) & -\alpha_k (s_q \beta_k - s_k \beta_q^*)
\end{pmatrix}
\begin{pmatrix}
\nu_q \\
\chi_q
\end{pmatrix}.
\end{align}

These expressions are cumbersome, but for a ground state calculation of a half-filled model we only need a single entry, as shown in Appendix \ref{app:pt-details}.

The perturbative expansion is:
$F_e \approx F_e^{(0)} + F_e^{(1)} + F_e^{(2)} + ...$
where the corrections are given by the usual quantum-mechanical expressions, namely
\begin{align}
F_e^{(1)} &= \bra{\Psi_0} \hat{U}_{el-ph} \ket{\Psi_0}, \\
F_e^{(2)} &= \bra{\Psi_0} \hat{U}_{el-ph} \frac{(1 - \ket{\Psi_0} \bra{\Psi_0})}{F_e^{(0)} - \hat{H}_{\text{unper}}} \hat{U}_{el-ph} \ket{\Psi_0}.
\end{align}
Here $\ket{\Psi_0}$ is the electronic Slater-determinant ground state of the half-filled SSH model, consisting of a full valence band and an empty conduction band. This leads to:
\begin{equation}\label{eq:delta_knm}
\delta K_{nm} = \frac{\partial^2 F_e^{(2)}}{\partial x_n \partial x_m} 
=  \bra{\Psi_0} \hat{f}_n \frac{(1 - \ket{\Psi_0} \bra{\Psi_0})}{E_0 - \hat{H}_{\text{unper}}}  \hat{f}_m \ket{\Psi_0} + \bra{\Psi_0} \hat{f}_m \frac{(1 - \ket{\Psi_0} \bra{\Psi_0})}{E_0 - \hat{H}_{\text{unper}}}  \hat{f}_n \ket{\Psi_0}
\end{equation}

Now we have to work our way through several cases depending on the even/odd character of $n,m$. The details are relegated to Appendix \ref{app:pt-details}. The main results are the expressions for the dynamical matrix entries $\delta K_{nm}$ listed below (the factor of 2 is from the sum over spins.)

\textbf{Case 1:} $n$ even, $m$ even, or $n$ odd, $m$ odd.

\begin{equation}\label{eq:e-e}
\delta K_{2n,2m} = \delta K_{2n+1, 2m+1} =  -2\left(\frac{4 \alpha}{N} \right)^2 \sum_{ |k,q|<\frac{\pi}{2a} } \left( \frac{\cos[ (k-q)(2n-2m)a ]}{E_k + E_q} \right) \left|s_k \beta_k + s_q \beta_q^*\right|^2
\end{equation}  

\textbf{Case 2:} $n$ even, $m$ odd.
\begin{equation}\label{eq:e-o}
\delta K_{2n,2m+1} =  +2\left(\frac{4  \alpha}{N} \right)^2 \sum_{ |k,q|<\frac{\pi}{2a} } \left( \frac{ e^{-i(k-q)(2m-2n+1)a}}{2(E_k + E_q)} \right) \left\{ (s_q \beta_q + s_k \beta_k^*) (s_k \beta_q^* + s_q \beta_k) \right\} + \text{h.c}.
\end{equation}

\textbf{Case 3:} $n$ odd, $m$ even.

\begin{equation}\label{eq:o-e}
\delta K_{2n+1,2m} =  +2\left(\frac{4  \alpha}{N} \right)^2 \sum_{ |k,q|<\frac{\pi}{2a} } \left( \frac{ e^{-i(k-q)(2m-2n-1)a}}{2(E_k + E_q)} \right) \left\{ (s_q \beta_q + s_k \beta_k^*) (s_k \beta_q^* + s_q \beta_k) \right\} + \text{h.c}.
\end{equation}

\end{widetext} 

Note that the sums over $k,q$ run over the reduced Brillouin zone $[-\pi/2a,\pi/2a]$ due to the dimerization ansatz.

While the expressions for the even-even and odd-odd sites turn out to be identical, there are subtle differences for the cross-terms that lift the phonon spectrum degeneracy at the Brillouin zone edge and split the optical phonon band from the acoustic band. In particular, there are two non-equivalent terms $\delta K_{2n,2n+1} \equiv Y_+, \delta K_{2n+1,2n+2} \equiv Y_-$, which correspond to the fact that there are even and odd electronic operators that give conduction and valence bands, $c^{(e)}, c^{(o)} \sim \chi, \nu$. Curiously, the spring constant corrections for longer-range distances larger than $1$ do not depend on whether the starting site is even or odd. This is not easy to see from Eqs. (\ref{eq:e-e})-(\ref{eq:o-e}), but once the integrals are carried out, the values obey $\delta K_{2n,2(n+\delta)-1} = \delta K_{2n+1,2(n+\delta)}$ for $\delta > 1$. If this were not the case, the unit cell for the phonon spectrum would be more than double: as it stands, it is the same size as the electronic unit cell. Moreover, it is remarkable that even though we started with only nearest-neighbour atomic force constants, through interactions with the extended electronic states we now have all-to-all force constants emerging. Because the Eqs. (\ref{eq:e-e})-(\ref{eq:o-e})  are closed-form expressions, we can easily calculate their values using computer integration, after going to the thermodynamic limit $\sum_k \rightarrow \frac{Na}{2\pi} \int dk$.

The next and final step is to diagonalize the ionic Hamiltonian with these new dynamical matrix elements, which we do in the next section.

\subsection{Finding the new phonon spectrum}

Since only the nearest-neighbour spring constant corrections differ depending on the bond, we can adopt the following notation for the dynamical matrix corrections: given arbitrary $n$, the same-site correction $Z_0 = 2K + \delta K_{n,n}$; two-site away correction $Z_2 = \delta K_{n,n+2}$; the even and odd bond corrections $Y_+ = -K+\delta K_{2n, 2n+1}, Y_- = -K+\delta K_{2n+1, 2n+2}$; and the remaining longer-range corrections $Z_{\delta \geq 1} = \delta K_{n, n+\delta}$.
Re-naming the operators as $\hat{p}_n, \hat{P}_n, \hat{x}_n, \hat{X}_n$ for the even and odd lattice sites of the $n^{th}$ unit cell, respectively  (our unit cell is taken to start at the even site) we rewrite:

\begin{widetext}
\begin{multline}\label{eq:ham_ion_final}
\hat{H}_{i,\text{har}}^{\text{(BO)}} = \sum_n  \frac{\hat{p}_{n}^2 + \hat{P}_n^2}{2M} 
+ \frac{1}{2} \sum_n \Big[ Z_0 (\hat{x}_n^2 + \hat{X}_n^2) + Y_+ \hat{x}_n \hat{X}_n + Y_- \hat{X}_n \hat{x}_{n+1} + Z_2 (\hat{x}_n \hat{x}_{n+1} + \hat{X}_n \hat{X}_{n+1} ) + \\
+ \sum_{\delta \geq 1} ( Z_{2+2\delta} \hat{x}_n \hat{x}_{n+1+\delta} + Z_{1+2\delta} 
\hat{x}_n \hat{X}_{n+\delta}
+ Z_{2+2\delta} \hat{X}_n \hat{X}_{n+1+\delta} + Z_{1+2\delta} \hat{X}_n \hat{x}_{n+\delta}  ) \Big] + E_{e}(\{R_n^0\}).
\end{multline}
\end{widetext}
Because $Y_+ \neq Y_-$ it is clear that the unit cell will double, and thus we have a resulting optical phonon branch (in the folded Brillouin zone), even though we started with only a bare acoustic branch (in the full Brillouin zone). The longer-range terms do  not affect the size of the unit cell, but do further renormalize the phonon spectrum. 

The resulting phonon spectrum (see Appendix \ref{app:ionic-diag} for details on the diagonalization of Eq. (\ref{eq:ham_ion_final})) is:
\begin{widetext}
\begin{multline}\label{eq:phon_spect}
\omega^2_{jq} = \frac{Z_0 - 2\sum_{\delta \geq 1} \left|Z_{2\delta}\right| \cos(2\delta qa) }{M} \pm \frac{1}{M} \Big\{ \Big[Y_+^2 + Y_-^2 + 2 Y_+ Y_- \cos(2qa)\Big] + \\
+ 4 \Big[\sum_{\delta \geq 1} Z_{1+2\delta} \cos((1+2\delta)qa) \Big]^2 + 8 (Y_+ + Y_-) \cos(qa) \sum_{\delta \geq 1} Z_{1+2\delta} \cos((1+2\delta)qa)  \Big\}^{1/2}.
\end{multline}
\end{widetext}
The expression for the phonon spectrum in Eq. (\ref{eq:phon_spect}) clearly demonstrates the appearance of an optical branch separated from the acoustic one if $\alpha\ne 0$. 
In the absence of electron-phonon coupling we of course have $Z_0 = 2K$, $Z_{\delta \geq 2} = 0, Y_{\pm} = -K$, from which we readily recover the undisturbed, folded acoustic spectrum
\begin{equation}\label{eq:undisturbed_phon}
    \omega_{jq} \xrightarrow{\alpha \rightarrow 0} \sqrt{ \frac{2K}{M} } \sqrt{ 1 + (-1)^j \cos(qa) } .
\end{equation}

For a finite electron-phonon coupling, the number of $Z_{\delta}$ to be included in the calculation, defined by the cut-off $\left| \delta \right| < \delta_{\text{max}}$, depends on how quickly they decay as a function of distance $\delta$ between unit cells. The appropriate value for $\delta_{\text{max}}$ is selected so as to insure the convergence of the phonon spectrum and ground state energies.

In the appropriate basis of phonon operators $b_{jq}, \hc{b}_{jq}$, we thus have:
\begin{equation}
    \hat{H}_{i,\text{har}}^{\text{(BO)}} = \sum_{jq} \hbar \omega_{jq} \left( \hc{b}_{jq} b_{jq} + \frac{1}{2} \right) + E_e(\{ R_n^0 \}).
\end{equation}

To complete the calculation and find the ground state, we need to find the value of $u$ that minimizes the total energy of the system. Furthermore, this value of $u$ must lead to a well-defined, real phonon spectrum $\omega_{jq}$. 
The total energy per site $E(u)$ of the system at $T = 0$ is:
\begin{equation}\label{eq:tot_energy}
    N E(u) = \sum_{jq} \frac{\hbar \omega_{jq} }{2}+ E_e(\{ R_n^0 \}).
\end{equation}
Because the renormalization of the phonon spectrum depends on the value of $u$, the zero-point energy (ZPE) of the lattice contributes to determining  the equilibrium value of $u$, unlike in the SSH approach where only the electronic contribution (second term) is considered. 

We minimize the energy in Eq. \ref{eq:tot_energy} numerically, thereby finding the ground state of the model. In the next section, we discuss the phonon spectrum renormalization in detail for the SSH-like model parameters. We also compare the phonon spectra obtained by our method with others from the literature for linear acetylenic chains (carbyne), study the validity of the $\Gamma$-point approximation for the phonon calculation, and consider how the finite size of the chain and the isotope effect affect the chain's zero-point energy, and what role they play in determining the ground state dimerization. In what follows, we will refer to our calculation as the BO+Har approach, and label it accordingly in the figures.

\section{Results and Discussion}\label{sec:res-and-disc}

\subsection{Comparison with the SSH calculation}
\label{subsec:cf_ssh}
First, we study the phonon spectrum of the model without any dimerization, $u = 0$ We adopt the canonical SSH parameter values $t = \SI{2.5}{\electronvolt}, \alpha = \SI{4.16}{\electronvolt / \angstrom}, K = \SI{21}{\electronvolt / \angstrom^2}$, with the lattice constant $a = \SI{1.22}{\angstrom}$ \cite{Su1979}. Following Ovchinnikov \textit{et al.}, for the mass of the C-H unit we use an appropriate ``reduced mass'' value of $M = \SI{2.16e-26}{\kilo\gram}$\cite{Ovchinnikov1982}. 

\begin{figure}
    \centering
    \includegraphics[width=0.9\linewidth]{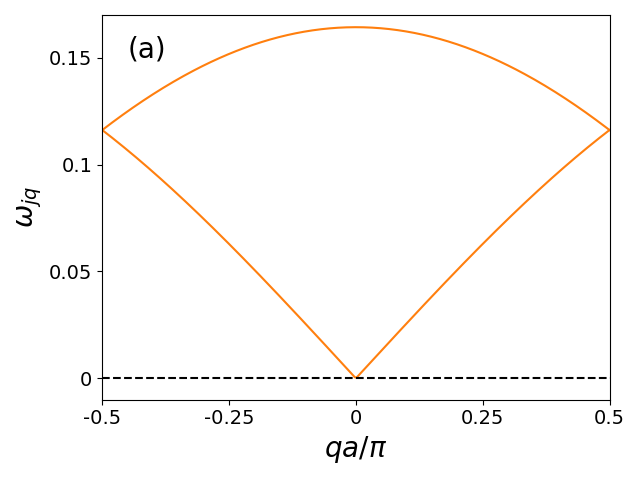}
    \includegraphics[width=0.9\linewidth]{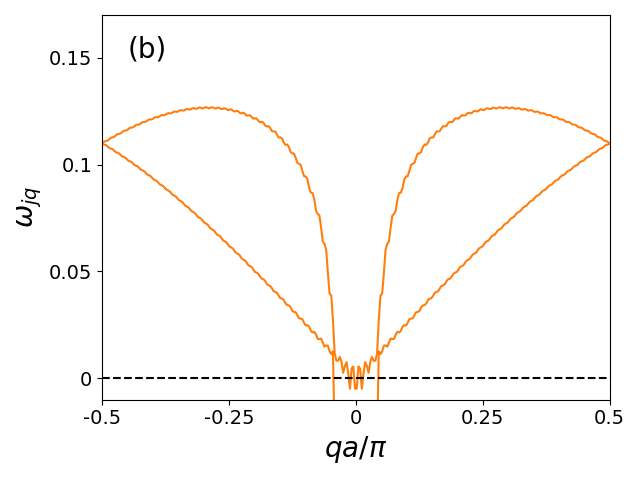}
    \includegraphics[width=0.9\linewidth]{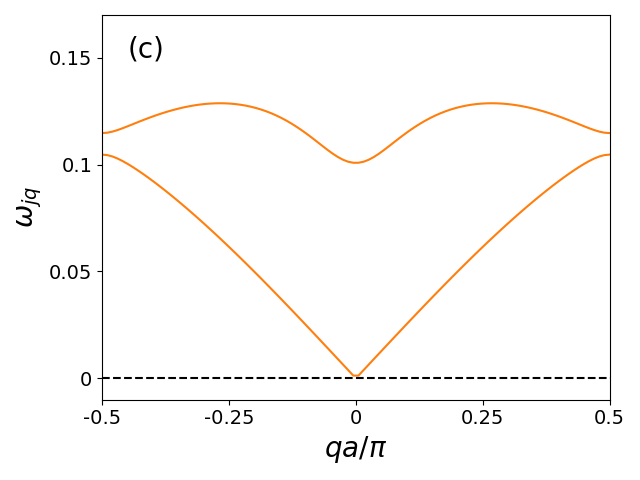}
    \caption{Phonon spectra $\omega_{jq}$ in eV (a) assuming no effect of the electrons on the phonon spectrum (standard SSH mean-field), (b) with our method but no dimerization, $u=0$; and (c) with our method at the dimerization $u = u_{\text{min}}$ that minimizes the total energy. Parameters are $t = \SI{2.5}{\electronvolt}, \alpha = \SI{4.16}{\electronvolt / \angstrom}, K = \SI{21}{\electronvolt / \angstrom^2}, a = \SI{1.22}{\angstrom}, M = \SI{2.16e-26}{\kilo\gram}$. The giant Kohn anomaly is clearly visible in the optical branch (the upper branch in the figure) at $q=0$ in panel (b) -- the anomaly is so large it leads to imaginary frequencies, plotted here as negative frequencies according to standard convention. It is lifted when the chain is allowed to dimerize, its only remnant being a slight softening of the optical branch at $q = 0$. The convergence parameter $\delta_{\text{max}} = 160$ for panels (b) and (c) (although convergence is achieved already at $\delta_{\text{max}} = 20$ for panel (c)).}
    \label{fig:phonon_kohn}
\end{figure}

\begin{figure}
    \centering
    \includegraphics[width=0.9\linewidth]{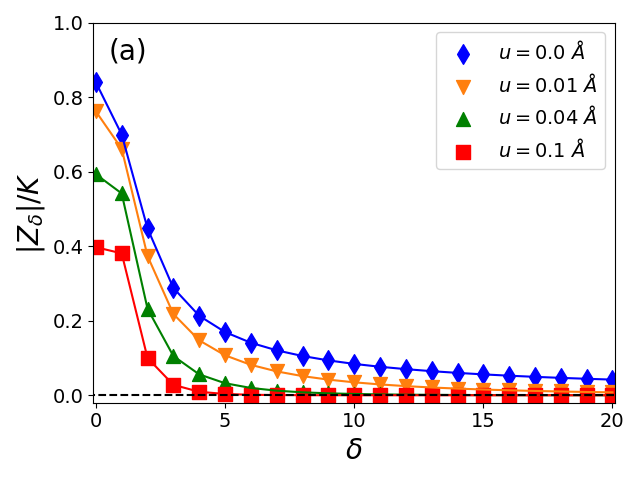}
    \includegraphics[width=0.9\linewidth]{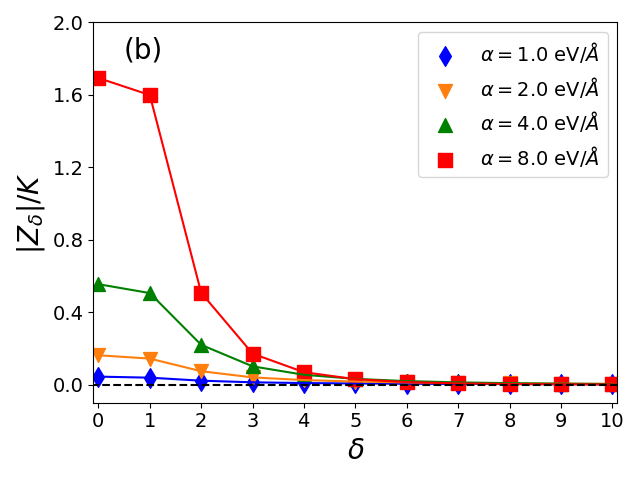}
    \caption{ Absolute value of force constants $|Z_{\delta}|$/K vs. $\delta$ (cf. Eq. \ref{eq:ham_ion_final}) (a) for the SSH-like coupling $\alpha = \SI{4.16}{\electronvolt \angstrom}$ and several values of $u$. In general, the closer to $u = 0$, the larger the force constants (the equilibrium distortion is $u \approx \SI{0.04}{\angstrom}$). In panel (b) we plot the force constants for several values of $\alpha$, with $u = \SI{0.04}{\angstrom}$ and otherwise SSH-like model parameter values: the stronger the coupling, the higher the force constants. The lines are guides to the eye. } 
    \label{fig:force_constants}
\end{figure}

The traditional mean-field calculation neglects any back-reaction of the electrons on the phonons and  predicts an  undisturbed, folded acoustic phonon band (Eq. \ref{eq:undisturbed_phon}), as shown in panel (a) of Fig. \ref{fig:phonon_kohn}. By contrast, our calculation finds a massive Kohn anomaly in the optical phonon branch at $q = 0$ (the folded $q = 2 k_F$ point) for the undimerized ($u=0$) chain, as expected for a system that is unstable to dimerization (see panel (b) of Fig. \ref{fig:phonon_kohn}). The anomaly is so strong that it results in the phonon spectrum becoming purely imaginary near $q = 0$: for such values we take their magnitude and plot them as negative, in accordance with the usual convention for unstable lattice calculations. The Kohn anomaly arises out of the many new force constants that appear due to the electron-phonon coupling, as shown in Fig. \ref{fig:force_constants}. This is a difficult limit for our calculation due to the strong singularity in the denominators of Eqs. \ref{eq:e-e}-\ref{eq:o-e} when $u = 0$. This makes the force constant corrections $Z_{\delta}$ decay very slowly with $\delta$, so that a $\delta_{\text{max}} \sim 160$ is needed for convergence (the biggest obstacle to convergence is the $q = 0$ point of the acoustic spectrum as, owing to its infinite-wavelength limit character, it gathers contributions from the farthest-reaching spring constants).  Aside from the anomaly, the rest of the spectrum is unaffected, and no gap opens between the acoustic and optical branch. The force constants generally decrease as the distortion moves away from the singularity at $u = 0$, as can be seen in panel (b) of Fig. \ref{fig:force_constants}: at the same time, increasing the electron-phonon coupling strength generally increases the magnitude of the force constants.

\begin{figure}
    \centering
    \includegraphics[width=1.\linewidth]{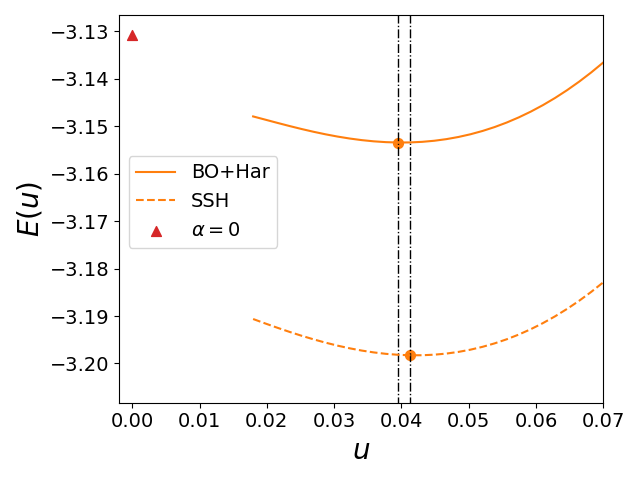}
    \caption{ Total energy per site $E(u)$ in eV versus chain dimerization $u$ in $\AA$, for our approach BO-Har (solid line), versus the mean-field SSH approach (dashed line). The dots indicate the corresponding ground states, with vertical dot-dash lines as guides to the eye. The red triangle is the total energy of the system in the absence of electron-phonon coupling. The convergence parameter is $\delta_{\text{max}} = 160$, the other parameters are as for previous figures. See text for more details.}
    \label{fig:total_energy_cf}
\end{figure}

For $u\ne 0$ the chain dimerizes doubling the unit cell, and our calculation shows that a proper optical phonon branch emerges, separated from the acoustic branch by a gap ($\sim \SI{5}{\milli\electronvolt}$) at $k = k_F$, see panel (c) of Fig. \ref{fig:phonon_kohn}. The Kohn anomaly is lifted, and only a remnant is left in the form of a softening of the $q=0$ optical phonon mode, of $\sim \SI{10}{\milli\electronvolt}$ for SSH-like values, relative to the un-dimerized chain. We note that here all phonon frequencies are real and positive, showing that the dimerized configuration represents a stable equilibrium. 

The total system energy as a function of $u$ is essentially modified in the vicinity of $u = 0$, due to the strong Kohn anomaly. The energies obtained by the standard SSH calculation (disregarding the renormalization of the phonon spectrum and its associated ZPE) and by the current BO+Har approach, which includes the ZPE, are compared in Fig. \ref{fig:total_energy_cf}. Near $u = 0$, the Kohn anomaly is so large that it leads to imaginary phonon energies for $u \lesssim \SI{0.01}{\angstrom}$; this is why the energies are only plotted for $u > \SI{0.02}{\angstrom}$ in Fig. \ref{fig:total_energy_cf}. For all shown $u$ values, there is a significant ZPE contribution to the total energy. While there is an overall upward shift of the energy, the ZPE contribution reduces the total energy for dimerization closer to $u = 0$, reducing the Peierls dimerization barrier and shifting the ground state dimerization value to $\sim \SI{0.039}{\angstrom}$, down from \SI{0.041}{\angstrom} without the ZPE contribution (amounting to a $\sim 5\%$ change -- a small, but noticeable effect). The vertical dot-dashed lines help illustrate this dimerization change.

Because the harmonic expansion is carried out to second order in displacements, we also did the similar calculation for the \textit{quadratic} SSH Hamiltonian, which has the additional quadratic electron-phonon interaction term $\beta (u_{n+1} - u_n)^2 (\hc{c}_{n\sigma} c_{n+1,\sigma} + \text{H.c.})$ in each $t_{n,n+1}$. The calculation details and resulting force constant expressions are given in Appendix \ref{app:quad-SSH-details}. Even though such terms should be included for consistency, their impact on the results presented in this paper was found to be insignificant, and so all the results in the paper are presented for $\beta = 0$. However, this assumes that the other parameters are set to polyacetylene-like values. The situation might be different for very different parameters.

\subsection{Comparison to other methods for polyacetylene}

\begin{figure}
    \centering
    \includegraphics[width=0.9\linewidth]{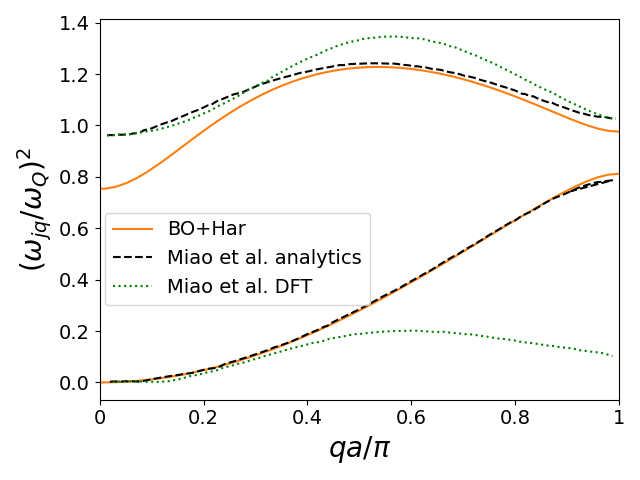}
    \caption{Comparison of scaled phonon spectra computed using our method (orange solid line) versus results from Miao \textit{et al.}\cite{Miao1999}: analytical (black dashed line) and DFT (dotted green line). The parameters we use are as in Fig. \ref{fig:total_energy_cf}, with convergence achieved with $\delta_{\text{max}} = 40$. See text for more details.} 
    \label{fig:cf_miao}
\end{figure}

It is interesting to compare the approach used here to other predictions for the phonon spectrum in the presence of electron-phonon coupling. In Fig. \ref{fig:cf_miao} we compare the phonon spectrum obtained for polyacetylene using the current method against results from DFT on 10-site supercells of polyacetylene by Miao \textit{ et al.} \cite{Miao1999}, as well as an earlier analytical calculation similar in spirit to the current approach by Fan \textit{et al.} \cite{Fan1988}. It is unclear to us what model parameters were used by the authors for generating the analytical results (chosen so as to fit the endpoints of the DFT-generated optical band). This is why we plot  the phonon spectra $\omega_{jq}$ scaled by the bare optical phonon frequency $\omega_Q = \sqrt{2K/M^*}$. For the study by Miao \textit{et al.}, we estimate their $\omega_Q^{\text{st}} = \SI{1420}{\centi\meter^{-1}}$ from the available results.

With this scale adjustment, we find reasonable agreement between our approach and that of Miao \textit{et al.}. The agreement for the optical phonon branch is great for all three methods, with our method producing a slightly larger phonon softening but smaller phonon gap. In the acoustic branch the agreement is perfect between the two analytical methods, while DFT  predicts a strongly flattened acoustic branch, similar to what is seen in other DFT studies of polyacetylene \cite{Lynge2004}. Miao \textit{et al.} hypothesize that this difference is because the SSH model treats only phonons along the chain axis, whereas there is significant coupling between the low-wavenumber acoustic mode and modes perpendicular to the chain axis, included in the DFT studies. More work is needed to clarify this point. 

\subsection{Comparison to other methods for carbyne}

Any use of this method to systems other than polycetylene must keep in mind the two main limitations: a) the validity of the harmonic approximation, and b) the applicability of the SSH model to the description of electron-phonon interactions. The latter can prevent us from accessing some regimes of interest, for instance the recently reported \textit{ab initio} study of the hypothetical 1D hydrogen chain \cite{Derriche2021}, where  $2 \alpha u \gg t$ places the system in the strongly non-linear coupling regime. 

Carbyne -- a pure carbon chain with alternating single and triple bonds -- is  convenient for such a comparison, given the similarly of its crystal structure to polyacetylene. However, carbyne has two degenerate electron orbitals, $p_y$ and $p_z$, that can host delocalized $\pi$-bond electrons, compared to polyacetylene's single $p_z$ orbital. This leads to an extra overall factor of 2 in Eqs. (\ref{eq:e-e})-(\ref{eq:o-e}).

In Ref. \onlinecite{Milani2008}, carbyne was studied with a semi-empirical bond-bond polarization approach to construct a force field that gave good agreement with experimental measurements of Raman spectra. By adjusting model parameters to match the observed electronic band gap and Raman excitation frequencies, the authors adopted $t = \SI{6.15}{\electronvolt}, \alpha = \SI{7.6}{\electronvolt/\angstrom}, K = \SI{81}{\electronvolt/\angstrom^2}, M = \SI{1.99e-26}{\kilo\gram}$ and found dimerization of $u = \SI{0.088}{\angstrom}$. 
However, their approach did not explicitly minimize the ground state energy against dimerization or other crystal lattice parameters, and thus does not provide a theoretical origin for the ground state dimerization. 

For these parameters, our method predicts that the ground state is undimerized: the spring constant is too stiff to allow any dimerization, as the semiclassical dimerization cost $\sim Ku^2$ grows very quickly. However, it is known from numerous previous experimental and \textit{ab initio} studies that carbyne (and finite-length polyynes) are indeed dimerized. The fact that our method predicts an undimerized ground state is possibly due to the exclusion of electron-electron interactions from the analysis, whose importance was recognized by Ovchinnikov long ago \cite{Ovchinnikov1982}. 

\begin{figure}
    \centering
    \includegraphics[width=0.9\linewidth]{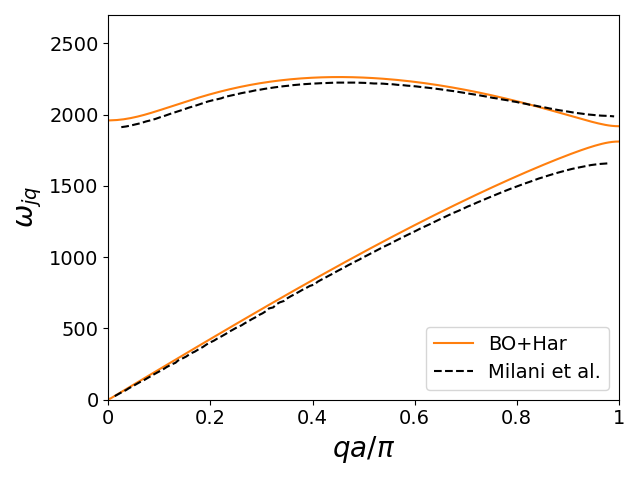}
    \caption{Phonon spectrum $\omega_{jq}$ of carbyne, in cm$^{-1}$, computed using our BO+Har method (orange solid line) versus that of Milani \textit{et al}\cite{Milani2008} (black dashed line). We use the same parameters $t = \SI{6.15}{\electronvolt}, \alpha = \SI{7.6}{\electronvolt/\angstrom}, K = \SI{81}{\electronvolt/\angstrom^2}, M = \SI{1.99e-26}{\kilo\gram}$ but set $u = \SI{0.044}{\angstrom}$ to reproduce the electronic band gap of $\SI{2.7}{\electronvolt}$ used in Ref. \onlinecite{Milani2008}. Convergence was achieved with $\delta_{\text{max}} = 40$.  } 
    \label{fig:cf_milani}
\end{figure}

To compare phonon spectra generated by the two approaches, we set $u = \SI{0.044}{\angstrom}$, so as to reproduce the electronic band gap of $\SI{2.7}{\electronvolt}$. The comparison, shown in Fig. \ref{fig:cf_milani}, is to the results reported in Fig. 3 of Ref. \onlinecite{Milani2008}.  We find strong agreement especially for the optical phonon branch, however the phonon gap is somewhat smaller in our approach. 

\subsection{The $\Gamma$-point approximation}

\begin{figure}[t]
    \centering
    \includegraphics[width=0.9\linewidth]{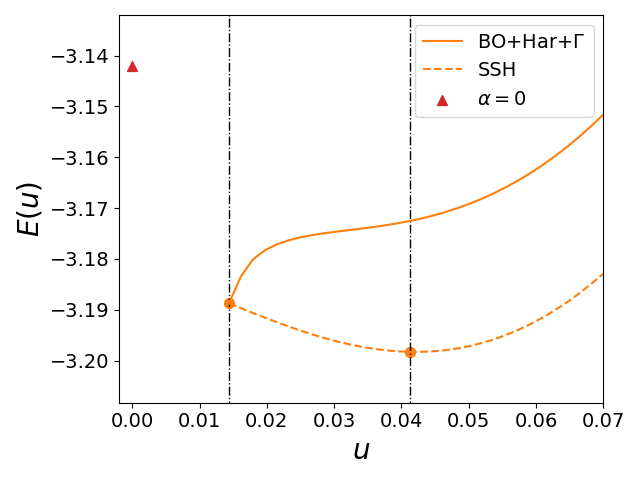}
    \caption{Same as Figure \ref{fig:total_energy_cf}, but the $\Gamma$-point approximation is used to generate the ZPE contribution (i.e., only the energy of the $q = 0$ phonon modes is counted in the BO+Har approach, scaled by the system size). This additional approximation predicts a significantly reduced ground state dimerization -- if it were not for the fact that the phonon spectrum becomes imaginary closer to $u = 0$ (energy not shown), the undimerized chain would be the predicted ground state. The two curves touch when the optical frequency vanishes at $q = 0$.}
    \label{fig:hudson_be_wrong}
\end{figure}

\begin{figure}[t]
    \centering
    \includegraphics[width=0.9\linewidth]{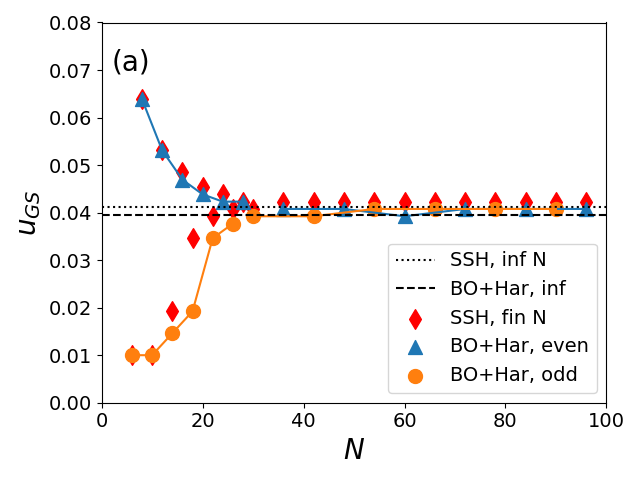}
    \includegraphics[width=0.9\linewidth]{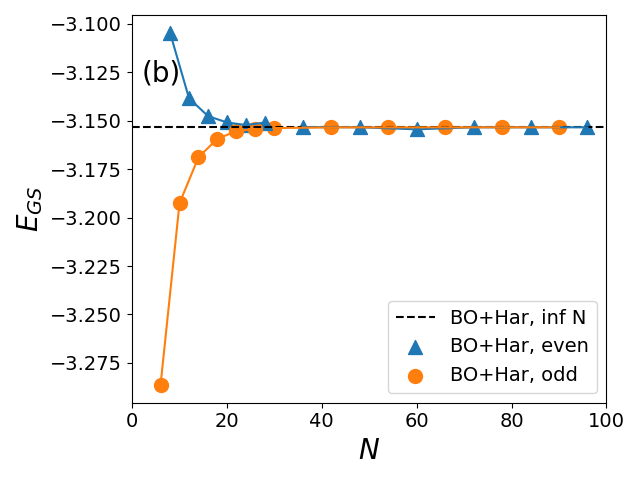}
    \caption{Ground-state dimerization $u_{\text{GS}}$ (panel (a)) and ground state energy $E_{\text{GS}}$ (panel (b)) for finite-$N$ chains with $N/2$ odd (orange circles) and even (blue triangles), as well as the SSH finite $N$ results (red diamonds in panel (a)). The lines connecting the circles and triangles are guides to the eye. The SSH and BO+Har infinite $N$ limits are given by the dotted and dashed lines, respectively, and are denoted ``SSH inf'' and ``BO+Har inf'' in the legend. The even/odd behaviour is explained by the momentum points that are allowed for each $N$, and whether they fall on the BZ edge or not. Finite $N$ results approach corresponding infinite $N$ results for both the SSH and the BO+Har approach. The effect of ZPE on ground state dimerization is visible in panel (a), where the SSH predicts a higher $u_{\text{GS}}$ than the BO+Har for $N>14$.}
    \label{fig:finite_n}
\end{figure}

From the renormalized phonon dispersion in panels (b) and (c) of Fig. \ref{fig:phonon_kohn}, we see that the modes most strongly affected by the electron-phonon coupling are those at and near $q = 0$; the rest of the spectrum is little affected. If only this $q=0$ mode is considered when searching for the equilibrium lattice structure, \textit{i.e.} a $\Gamma$-point approximation is employed\cite{Hudson2013}, the effect of the ZPE will be significantly overestimated. To exemplify this, in Fig. \ref{fig:hudson_be_wrong} we repeat Fig. \ref{fig:total_energy_cf} for SSH-like values, but assuming that the whole ZPE  comes from the $q = 0$ modes (scaled by system size $N/2$ for a proper comparison). Clearly, this additional approximation significantly changes the results, and in particular it predicts a ground state with significantly reduced dimerization (a 65\% reduction relative to the standard SSH result). If it were not for the fact that the phonon spectrum becomes unstable near $u =0$, that would be the predicted ground state -- as indeed found in Ref. \onlinecite{Hudson2013}, where the $\Gamma$-point approximation was employed. We hope that this comparison goes some way to address the controversy whether there is dimerization for an infinite polyacetylene-like chain, and emphasizes the importance of using the entire phonon spectrum in selecting the true ground state. 

\subsection{Finite length chains}

If we calculate the discrete sums in Eqs. (\ref{eq:e-e}-\ref{eq:o-e}) instead of taking the thermodynamic limit $N\rightarrow \infty$, we can study finite-size effects for chains with an even number of atoms. Now we fix $\delta = N/2$ where $N$ is the total number of atoms, since the equations include spring constants to $+\delta$ and $-\delta$ sites. The physical point of comparison here is something like a benzene ring ($N = 6$) and longer periodic ring structures. Cyclic polyenes are known to have a transition from all-equal (undimerized) to conjugated bonds at some critical size, typically pegged at $N \sim 8$. In what follows, we always take the lowest available energy at the ground state energy, subject to the phonon spectrum being stable. 

In Fig. \ref{fig:finite_n} we plot the ground state dimerization $u_{\text{GS}}$ (panel (a)), and ground state energy $E_{\text{GS}}$ (panel (b)) as a function of chain length $N$ using the polyacetylene model parameters, calculated using the SSH and BO+Har approaches. For comparison, the corresponding infinite $N$ limits are given by the dotted and dashed lines, respectively, and are denoted ``SSH inf'' and ``BO+Har inf'' in the legend. As expected, there is a difference between the dimerization (panel (a)) predicted by the infinite $N$ SSH and BO+Har approaches (cf. Ref. \ref{fig:total_energy_cf}). We split our dataset into even and odd groups for plotting, where chains whose length $N$ is such that $N/2$ is even are in the even group (blue triangles), and the others are in the odd group (orange circles). We also show the SSH finite $N$ results to look at the change in the ground state dimerization driven by the ZPE (they also alternate between even/odd but we do not show this for clarity). The ground state energy differs little between the BO+Har and the SSH approach due to the flatness of the energy curve near the minimum other than the constant ZPE contribution -- for this reason the SSH energy is not shown in panel (b).

Convergence to the infinite $N$ results for both the energy and the ground state dimerization is achieved quickly, already by $N \sim 40$. The SSH finite $N$ dimerizations approach the SSH infinite $N$ limit, and similarly for the BO+Har approach. 

Curiously, even group BO+Har chains converge to infinite $N$ BO+Har results from above (bigger dimerization, higher energy at small $N$), while odd group chains approach from below (smaller dimerization, lower energy at small $N$). Fundamentally, this is a finite size effect: the alternation with increasing $N$ arises because the allowed momentum points are different in the two cases. In the even group, a pair of momentum points (one point for each of optical / acoustic, and valence / conduction bands) at the edge of the Brillouin zone is included: in the odd group, it is not. The impact of dimerization on the electronic bands (and thus on the total energy) is strongest at the edge of the Brillouin zone. When rings are short and only a few momentum points are allowed, there is a big difference if the edge is included or not. Once more points are allowed for longer chains, the alternation is strongly reduced. This alternation is seen in both the SSH and our approach -- it is not ZPE driven.

The effect of ZPE leads to stronger deviations from the SSH finite $N$ result at low $N$ for the chain dimerization than in the infinite $N$ limit: the BO+Har predicts a significantly smaller dimerization for $14 \lesssim N \lesssim 25$ for both even and odd groups than the SSH approach. This means the zero-point energy destabilizes the dimerization of the chain more effectively in the small $N$ limit. We are led to conclude that for smaller systems ($N \lesssim 20$), the effect of ZPE is significant in determining the true ground state of the system and a fully quantum-mechanical treatment is necessary to elucidate the extent of that effect. 

Using a smaller mass parameter is another way to increase the impact of the ZPE on the ground state dimerization, as we demonstrate in the next section. 

\subsection{Isotope effect}

\begin{figure}[t]
    \centering
    \includegraphics[width=0.8\linewidth]{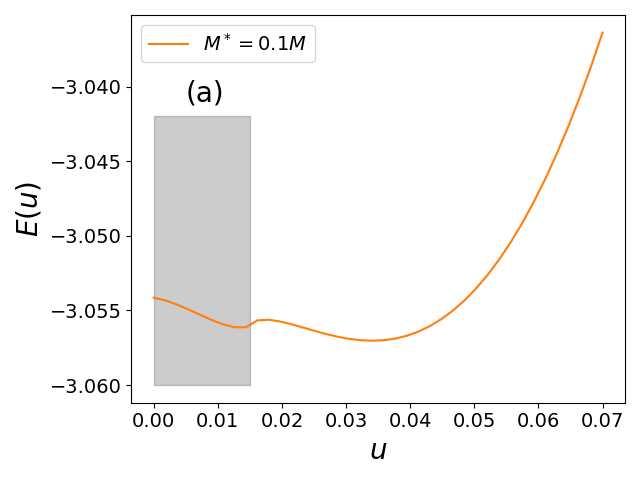}
    \includegraphics[width=0.8\linewidth]{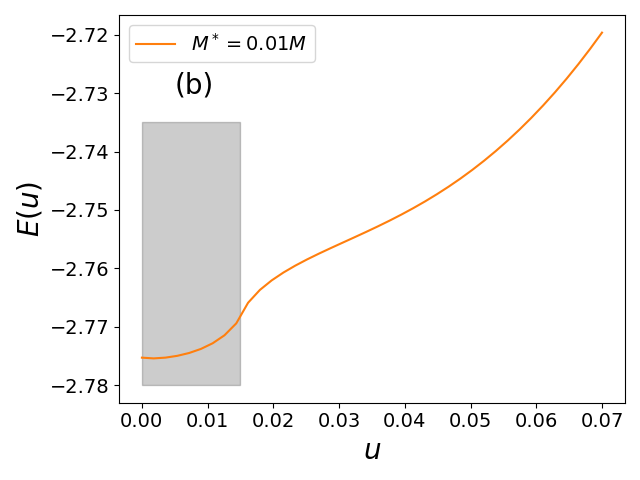}
    \includegraphics[width=0.8\linewidth]{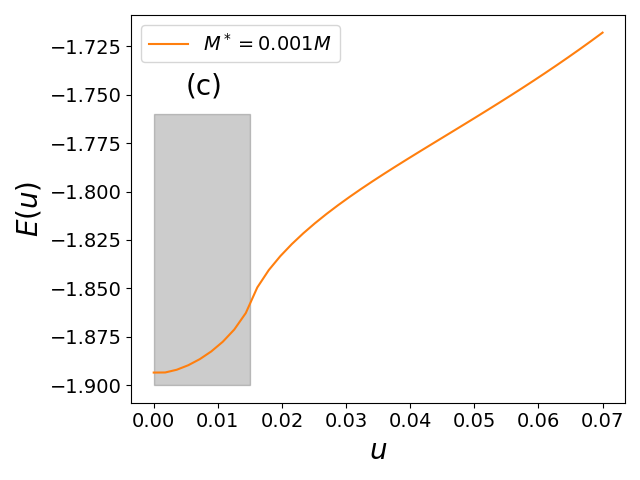}
    \caption{$E(u)$ versus $u$ for the  mass (a) $M^* = 0.1M$, (b) $M^* = 0.01M$, (c) $M^* = 0.001M$; all other parameters are the same as for polyacetylene. A significant reduction of the ionic mass boosts the characteristic phonon energy, and thus the importance of ZPE. A significantly smaller lattice dimerization is favored once $M^*$ becomes sufficiently small: it would be zero were it not for the phonon spectrum becoming unstable (the shaded region indicates the extent of $u$ such that the phonon spectrum is unstable). }
    \label{fig:small_mass}
\end{figure}

The results presented so far might suggest that calculating the renormalization of the phonon spectrum and including the ZPE in the total energy is an unwarranted complication, given that the end results are little affected. This is to be expected for polyacetylene, given the order of magnitude difference between the electronic and lattice energy scales. However, there might be contexts where the energy scales are not so disparate -- for instance, in a hypothetical hydrogen chain, or if the electron mass is strongly renormalized through interactions.

To exemplify this point, we lower the mass parameter $M$ as a simple way to boost the lattice energy scale. The results are shown in Fig. \ref{fig:small_mass}, where we compare the total energy curves for cases where the ionic mass $M^* = 0.1, 0.01, 0.001 M$,  $M$ being the polyacetylene value; the other parameters are kept unchanged. Indeed, when $M^*$ decreases by two orders of magnitude (so that the characteristic phonon energy increases by one order of magnitude and becomes comparable to the electronic energy scale), the addition of the ZPE predicts a much less dimerized GS. We caution against a literal interpretation of the energies close to $u = 0$, where the phonon spectrum is unstable  (the shaded region): however, we see that the previously existing potential well has been raised to the point where the energy goes up almost linearly, favouring smaller and smaller dimerization. Another caveat is that we explicitly used the fact that the electron masses are so much smaller than those of the lattice ions: as $M^*$ grows smaller, that approximation is called into question. For contrast, we note that the prediction of the standard SSH calculation (which does not include the renormalized ZPE) remains the same as the dashed line in Fig. \ref{fig:total_energy_cf} irrespective of the value of $M^*$. 

Of course, these latter results are artificial, as at most $M^* \approx 0.07M \approx m_p$ (for a hypothetical hydrogen 1D chain). However, the phonon energy scale can be increased not just by lowering $M$, but also by stiffening the spring constant $K$. The important point is that if $\sqrt{K/M}\sim t$, the ZPE contribution cannot be ignored and that it is likely strongly renormalized as a function of the dimerization $u$.

\section{Conclusions}\label{sec:concl}

In this paper, we used  the Born-Oppenheimer approximation together with a harmonic approximation to develop a relatively simple technique for calculating the phonon spectrum renormalization due to the electron-phonon coupling. In contrast to variational techniques such as SSCHA \cite{Monacelli2021} or SCHA \cite{Souvatzis2008}, the force constants are calculated directly instead of being assumed to be variational parameters adjusted to minimize the energy -- only the static lattice dimerization $u$ is treated this way. 

We applied this method to the SSH model with polyacetylene parameters, finding that a strong Kohn anomaly appears in the optical branch at $q = 0$ for an undimerized chain with $u=0$, signalling its instability to dimerization. This anomaly is replaced by a softening of the $\Gamma$-point optical phonon when the chain is allowed to assume its minimum energy (dimerized) state. Our phonon spectra for polyacetylene and carbyne show solid agreement with prior literature results, both analytical and \textit{ab initio}. We note the appearance of long-range force constants due to the electron-phonon interaction, even though the bare ion-ion interactions are nearest-neighbor only. Crucially, we find that the zero-point phonon energy contribution is unable to destroy the dimerized ground state of polyacetylene for an infinite chain. This agrees with several previous results and disagrees with others. Relevant to the latter,  we point out the potential downfall of using the $\Gamma$-point approximation, as it overestimates the ZPE dependence on $u$. Finally, we show that the role of the ZPE is amplified for finite-length chains, where the ground state ceases to be dimerized for $N = 6, 10$; or through lowering the mass parameter of the chain  and/or increasing its stiffness to the point where the characteristic electronic and ionic energy scales are more comparable.

The most significant approximations we made is to treat the ionic BO energy within the harmonic approximation. The force constants can be then calculated using perturbation theory, and an analytical expression for the phonon spectrum is available. In principle one can relax this approximation by including cubic and higher order unharmonic terms, whose coefficients can be calculated with the appropriate higher-order perturbation theory. However, additional approximations (eg., mean-field) are needed to then deal with these unharmonic terms. Given its small energy scale, we believe that for polyacetylene such a calculation is not warranted, but it might be relevant for a system with very different parameters. We note that we have tested the validity of the other approximation, namely of including only linear SSH electron-phonon coupling, by adding the next term and investigating its effects. Indeed, these were find to be negligible for polyacetylene parameter values. 

\begin{acknowledgments}
We are grateful to Alberto Nocera, George Sawatzky and Nassim Derriche for valuable discussions about the model and the use of Born-Oppenheimer approximation. This work was supported by the UBC Stewart Blusson Quantum Matter Institute,  the Max-Planck-UBC-UTokyo Center for Quantum Materials and the Natural Sciences and Engineering Research Council of Canada. We are also grateful for the use of computational resources at the LISA computational cluster at the Stewart Blusson Quantum Matter Institute.
\end{acknowledgments}

\appendix

\section{Perturbative calculation details}
\label{app:pt-details}
In this appendix we describe the details of getting from the general perturbative expressions in Eq. (\ref{eq:delta_knm}) to the results reported for $\delta K_{nm}$ in Eqs. \ref{eq:e-e}-\ref{eq:o-e}. First, we catalog the ``halves'' of the perturbative expressions for the even-odd cases.

\textbf{Even $m$}: 

\begin{multline}
\frac{(1 - \ket{\Psi_0} \bra{\Psi_0})}{E_0 - \hat{H}_{\text{unper}}}  \frac{\partial \hat{V}}{\partial x_{2m}} \ket{\Psi_0} = \\
= \frac{2i\alpha}{N/2} \sum_{kq\sigma} \frac{e^{-i(k-q)(2m)a}}{E_k + E_q} \alpha_k (s_k \beta_k + s_q \beta^*_q) \hc{\chi}_k \nu_q \ket{\Psi_0}.
\end{multline}

We arrived at this expression as follows: since on the right we have the ground state $\ket{\Psi_0}$, the only term from the matrix product in Eq. \ref{eq:V-even} that can survive at half-filling is the $\hc{\chi}\nu$ combination. As for the energy denominator, 
\begin{multline}
\left(E_0 - \hat{H}_{\text{unper}} \right) \hc{\chi}_k \nu_q \ket{\Psi_0} = \\
= \left( - \sum_p E_p \right) - \left[ \left( -\sum_p E_p \right) - (- E_k) +  (+ E_q) \right] = \\
= - (E_k + E_q).
\end{multline}
Given the reasoning above, we can immediately write down the odd version of the term

\textbf{Odd $m$}: 
\begin{multline}
\frac{(1 - \ket{\Psi_0} \bra{\Psi_0})}{E_0 - \hat{H}_{\text{unper}}}  \frac{\partial \hat{V}}{\partial x_{2m+1}} \ket{\Psi_0} = \\
= -\frac{2i\alpha}{N/2} \sum_{kq\sigma} \frac{e^{-i(k-q)(2m+1)a}}{E_k + E_q} \alpha_k (s_q \beta_k + s_k \beta^*_q) \times \\
\times \hc{\chi}_k \nu_q \ket{\Psi_0}.
\end{multline}

The corresponding left-hand sides of the overall perturbation expression are no different: now we keep the $\hc{\nu} \chi$ combination, so that it does not annihilate the state on the left

\textbf{Even $m$}: 
\begin{multline}
\bra{\Psi_0} \frac{\partial \hat{V}}{\partial x_{2m}} = -\frac{2i\alpha}{N/2} \sum_{k'q'\sigma}  e^{-i(k'-q')(2m)a} \alpha_k  \times \\
\times (s_{k'} \beta_{k'} + s_{q'} \beta^*_{q'}) \bra{\Psi_0} \hc{\nu}_{k'} \chi_{q'}.
\end{multline}

Connecting the left and right halves of the expressions forces $q' = k, k' = q$. Re-writing it for clarity

\begin{multline}
\bra{\Psi_0} \frac{\partial \hat{V}}{\partial x_{2m}} = -\frac{2i\alpha}{N/2} \sum_{kq\sigma}  e^{+i(k-q)(2m)a} \alpha_k \times \\ \times (s_{q} \beta_{q} + s_{k} \beta^*_{k}) \bra{\Psi_0} \hc{\nu}_{q} \chi_{k}.
\end{multline}

The only thing that changes for odd $m$ is the prefactor:

\textbf{Odd $m$}: 

\begin{multline}
\bra{\Psi_0} \frac{\partial \hat{V}}{\partial x_{2m+1}} = \frac{2i\alpha}{N/2} \sum_{kq\sigma}  e^{+i(k-q)(2m+1)a} \alpha_k \times \\ 
\times (s_{k} \beta_{q} + s_{q} \beta^*_{k}) \bra{\Psi_0} \hc{\nu}_{q} \chi_{k}.
\end{multline}

Now that we have all of the relevant expressions, we can begin to put them together. The first of Eqs. \ref{eq:e-e}-\ref{eq:o-e} is derived as follows

\begin{multline}
\delta K_{nm} = \left(\frac{2 \alpha}{N} \right)^2  \sum_{kq\sigma} \left( \frac{-e^{-i(k-q)(n-m)a}}{E_k + E_q} \right) \times \\
\times (4\alpha_k^2) \left\{ (s_k \beta_k + s_q \beta_q^*) (s_q \beta_q + s_k \beta_k^*) \right\} + (n\rightleftarrows m) = \\
= -2\left(\frac{4 \alpha}{N} \right)^2  \sum_{kq} \left( \frac{\cos[ (k-q)(n-m)a ]}{E_k + E_q} \right) \times \\
\times \left|s_k \beta_k + s_q \beta_q^*\right|^2
\end{multline} 
The other equations are obtained similarly.

\section{Extension to the quadratic SSH model}
\label{app:quad-SSH-details}

We  now use:
\begin{multline}
t(R_{n+1} - R_n) = t(a + u_{n+1} - u_n) \approx  \\
\approx t - \alpha_n (u_{n+1} - u_n) +  \frac{1}{2}\beta_{n} (u_{n+1} - u_n)^2.
\end{multline}
Here $\alpha_n$ and $\beta_{n}$ are first and second derivatives of the hopping integral with respect to $R_{n+1}-R_{n}$. The minus sign in front of $\alpha$ is convention to ensure $\alpha$ positive in magnitude. 

The usual expansion is only to first order. However, in the harmonic approximation of the effective ionic Hamiltonian,  the ion-ion interaction potential $\hat{V}(\hat{R}_{n+1} - \hat{R}_n)$ was expanded to second order, which makes a first order expansion for the hopping to appear inconsistent. We discuss here the consequences of using a quadratic Peierls expansions, which to our knowledge was not considered elsewhere.

Assuming constant values for $\alpha$ and  $\beta$ and defining $T_{n,n+1} = \sum_{\sigma} \hc{c}_{n+1,\sigma} c_{n\sigma} + \text{h.c.}$, we now find:
\begin{multline}
H_e = -(t+  2u^2\beta) \sum_{n} T_{n,n+1} - 2\alpha u \sum_{n} (-1)^n T_{n,n+1} + \\
+  \sum_{n} (\alpha + 2 \beta (-1)^n u) (x_{n+1} - x_n) T_{n,n+1} - \\
- \sum_{n} \frac{\beta}{2} (x_{n+1} - x_n)^2 T_{n,n+1} + E_e(\{R_n^0\}),
\end{multline}

The ``easy'' part can be diagonalized as before (the only change being $t \rightarrow t + 2 \beta u^2$), giving

\begin{multline}
H_e = - 2\sum_{k} E_k \left( \hc{\nu}_{k\sigma} \nu_{k\sigma} - \hc{\chi}_{k\sigma} \chi_{k\sigma} \right) + \\
+ \sum_{n} (\alpha + (-1)^n 2 \beta u) (x_{n+1} - x_n)  T_{n,n+1} - \\
- \sum_{n}\frac{\beta}{2}  (x_{n+1} - x_n)^2 T_{n,n+1} +  E_e(\{R_n^0\}).
\end{multline}

In the rest of this section, using results from Appendix \ref{app:pt-details}, we do the perturbative calculation for this second-order SSH model. Start with the ionic Hamiltonian from Eq. \ref{eq:ham_ion_final} (with the understanding that here $\epsilon_k = -2(t+2\beta u^2)\cos(ka)$)
\begin{multline}\label{eq:ion2-ham-pre-PT2}
\hat{H}_{i,\text{har}}^{\text{(BO)}} = \sum_n \frac{\hat{p}_n^2}{2M} + \frac{K}{2} \sum_{n} (\hat{x}_{n+1} - \hat{x}_n)^2 + \\
+ \frac{1}{2} \sum_{nm} \frac{\partial^2 \langle \hat{H}_e - \hat{V}_{i-i} \rangle}{\partial R_n \partial R_m} \hat{x}_n \hat{x}_m - \\
- 2 \frac{Na}{2\pi} \int_{-\frac{\pi}{2a}}^{\frac{\pi}{2a}} \, dk \sqrt{ \epsilon_k^2 + u^2\Delta_k^2 } + 2 K u^2 N.
\end{multline}
 The inter-ionic potential in this Hamiltonian is generated by various corrections up to second order to the energy of the electronic Hamiltonian, which reads
\begin{multline}\label{eq:el-ham2-pre-PT2} 
\hat{H}_{e} - \hat{V}_{i-i} = - \sum_{k\sigma} E_k ( \hc{\nu}_{k\sigma} \nu_{k\sigma} - \hc{\chi}_{k\sigma} \chi_{k\sigma} ) + \\
+ \underbrace{ \sum_{n\sigma} (\alpha + (-1)^n 2\beta u) (x_{n+1} - x_n) T_{n,n+1} }_{\equiv \hat{A}_{el-ph}} - \\
- \underbrace{ \sum_{n\sigma} \frac{\beta}{2} (x_{n+1} - x_n)^2 T_{n,n+1} }_{\hat{B}_{el-ph}}.
\end{multline}
All the definitions are as in Appendix \ref{app:pt-details}, with the caveat that $t \rightarrow t + 2\beta u^2$.

As before, we are evaluating energy corrections
\begin{equation}
F_e \approx F_e^{(0)} + F_e^{(1)} + F_e^{(2)} + ...
\end{equation}

where the terms are given by

\begin{align}
F_e^{(1)} &= \bra{\Psi_0} (\hat{A}_{el-ph} + \hat{B}_{el-ph}) \ket{\Psi_0}, \\
F_e^{(2)} &= \bra{\Psi_0} (\hat{A}_{el-ph} +\hat{B}_{el-ph} ) \frac{(1 - \ket{\Psi_0} \bra{\Psi_0})}{E_0 - \hat{H}_{\text{unper}}} \times \nonumber \\
&\times (\hat{A}_{el-ph} + \hat{B}_{el-ph} ) \ket{\Psi_0}.
\end{align}

Introduce the notation
\begin{multline}
\delta K_{nm} \equiv \frac{\partial^2 F_e}{\partial x_n \partial x_m} = 0 + \frac{\partial^2 F_e^{(1)}}{\partial x_n \partial x_m} + \frac{\partial^2 F_e^{(2)}}{\partial x_n \partial x_m} \equiv \\
\equiv \delta K_{nm}^{(1)} + \delta K_{nm}^{(2)}.
\end{multline}

Unlike in the conventional SSH model, the linear term from the perturbative expansion will have a non-zero contribution due to $\hat{B}_{el-ph}$ in the Hamiltonian. That is the new calculation: the second-order term will be simply appended with the new coefficients $\alpha_{\pm} = \alpha \pm 2\beta u$. There is no contribution from $\hat{B}_{el-ph}$ at second order because those will lead to terms cubic or quartic in atomic displacements, which will be set to zero in the harmonic approximation for the Born-Oppenheimer energy surface.

Start with the first-order term coming from $\hat{B}_{el-ph}$. For $\left| n-m \right| > 1$, the first-order correction is zero, as is clear from the following calculation:

\begin{multline}
\delta K_{nm}^{(1)} = \frac{\partial^2}{\partial x_n \partial _m} \left( \Big\langle - \frac{\beta}{2} \sum_{l} (x_{l+1} - x_l)^2 T_{l,l+1} \Big\rangle_{\Psi_0} \right) =  \\
= \frac{\partial^2}{\partial x_n \partial _m} \Big( \Big\langle - \beta \sum_{l}  \Big[ x_{l}^2 \left(T_{l,l+1} + T_{l-1,l}  \right) - \\
- x_l x_{l+1} T_{l,l+1} \Big] \Big\rangle_{\Psi_0} \Big) = \\
= -2 \delta_{nm} \beta \langle T_{n,n+1} + T_{n-1,n}\rangle_{\Psi_0} + \delta_{n,m+1} \beta \langle T_{n,n+1} \rangle_{\Psi_0}.
\end{multline}

Then:
\begin{multline}\label{eq:contr-quad-term-inter}
\delta K_{2n,2n}^{(1)} 
= -2\beta  \Big\langle T_{2n,2n+1} + T_{2n-1,2n} \Big\rangle_{\Psi_0}  \\
= -\frac{4\beta}{N/2} \sum_{kq\sigma} e^{-i(k-q)(2n)a} \times \\
\times \left[ \cos(ka) \langle \hc{c^{(o)}}_{k\sigma} c_{q\sigma}^{(e)} \rangle_{\Psi_0} + \cos(qa) \langle \hc{c^{(e)}}_{k\sigma} c_{q\sigma}^{(o)} \rangle_{\Psi_0} \right].
\end{multline}
To finish the evaluation, we must compute the expectation values above. To do so, we should express the old $c$ operators in terms of $\nu, \chi$ (at half-filling, we will take the un-perturbed ground state to be the usual Fermi sea $\ket{\Psi_0} = \prod_{ |k| <\pi/a,\sigma} \hc{\nu}_{k\sigma} \ket{0}$)

\begin{multline}
\langle \hc{c^{(o)}}_{k\sigma} c_{q\sigma}^{(e)} \rangle_{\Psi_0} = \langle (\beta_k \hc{\nu}_{k\sigma} - \beta_k \hc{\chi}_{k\sigma})(\alpha_q \nu_{q\sigma} + \alpha_q \chi_{q\sigma}) \rangle_{\Psi_0} = \\
= \delta_{kq} \beta_k \alpha_q.
\end{multline}
Similarly, for the other expectation value, we find at half-filling
\begin{multline}
\langle \hc{c^{(e)}}_{k\sigma} c_{q\sigma}^{(o)} \rangle_{\Psi_0} = \langle (\alpha_k \hc{\nu}_{k\sigma} + \alpha_k \hc{\chi}_{k\sigma})(\beta^*_q \nu_{q\sigma} - \beta^*_q \chi_{q\sigma}) \rangle_{\Psi_0} = \\
= \delta_{kq} \beta_q^* \alpha_k.
\end{multline}
Substituting into Eq. \ref{eq:contr-quad-term-inter}, we find
\begin{multline}
\delta K_{2n,2n}^{(1)} = \\
= - \frac{4\beta}{N/2} \sum_{kq\sigma} \delta_{kq} \left[ \cos(ka)\beta_k \alpha_q + \cos(qa) \beta_q^* \alpha_k \right] = \\
= - 2\frac{4\beta}{N/2} \sum_{k} \frac{\epsilon_k \cos(ka)}{E_k}.
\end{multline}

If $n$ is odd, the matrix elements swap places, but the end result remains
\begin{equation}\label{eq:contr-quad-term-inter2}
\delta K_{2n+1,2n+1}^{(1)} = - 2\frac{4\beta}{N/2} \sum_k \frac{\epsilon_k\cos(ka)}{E_k}.
\end{equation}
We may combine the two terms by writing
\begin{equation}
\delta K_{nn}^{(1)} = - 2\frac{4\beta}{N/2} \sum_k \frac{\epsilon_k \cos(ka)}{E_k}, \quad n \in \mathbb{Z}.
\end{equation}
It makes sense that the terms are the same for even and odd sites, as these are on-site corrections, and any site connects to a pair of short/long bonds.

After more similar calculations, we find
\begin{multline}
\delta K_{nm}^{(1)} = \delta_{nm} \left( - \frac{8\beta}{N/2} \sum_k \frac{\epsilon_k \cos(ka)}{E_k} \right) + \\
+ \delta_{n+1,m} \left( \frac{2\beta}{N/2} \sum_k \frac{\epsilon_k \cos(ka) + (-1)^nu\Delta_k \sin(ka)}{E_k} \right).
\end{multline}
Evidently, there are two inequivalent spring constants, and thus an optical and an acoustic branch will emerge for a nonzero $\alpha$.

Now for the contribution from $\hat{A}_{el-ph}$. As in the linear SSH model case, re-write it in terms of $\nu$ and $\chi$ operators,

\begin{equation}
\hat{A}_{el-ph}  \equiv \sum_{n} x_n \hat{g}_{n}
\end{equation}
where we defined $\alpha_{\pm}(n) = \alpha \pm (-1)^n 2\beta u$, and $\hat{g}_n =  -\alpha_+(n) T_{n,n+1} + \alpha_-(n) T_{n,n-1}$. Then  $\alpha_{\pm}(2n) = \alpha_{\pm}$, and 
\begin{multline}\label{eq:V-even2}
\hat{g}_{2n} 
=   \frac{ 1}{N/2} \sum_{kq\sigma} x_{2n} e^{-i(k-q)2na} \times \\
\times \Big[ (-2i \alpha \sin(qa) - 4\beta u\cos(qa)) \hc{c^{(e)}_{k\sigma}} c^{(o)}_{q\sigma} + \\
+ (2i \alpha \sin(ka) - 4 \beta u \cos(ka)) \hc{c_{k\sigma}^{(o)}} c_{q\sigma}^{(e)} \Big].
\end{multline}
while
\begin{multline}\label{eq:V-odd2}
\hat{g}_{2n+1} =  \frac{ 1}{N/2} \sum_{kq\sigma} x_{2n+1} e^{-i(k-q)(2n+1)a} \times \\
\times \Big[ (2i \alpha \sin(ka) + 4\beta u\cos(ka)) \hc{c^{(e)}_{k\sigma}} c^{(o)}_{q\sigma} + \\
+ (-2i \alpha \sin(qa) + 4 \beta u \cos(qa)) \hc{c_{k\sigma}^{(o)}} c_{q\sigma}^{(e)} \Big].
\end{multline}

The final step is to convert the operators to the $\nu, \chi$ basis. Defining $g_k = 2i \alpha\sin(ka) - 4 \beta u \cos(ka)$ and $h_k = 2i\alpha \sin(ka) + 4 \beta u \cos(ka)$, we find

\begin{widetext}
\begin{align}
\hat{g}_{2n} &= \frac{ 1}{N/2} \sum_{kq\sigma} x_{2n} e^{-i(k-q)(2n)a} \begin{pmatrix}
\hc{\nu}_k & \hc{\chi}_k
\end{pmatrix}
\begin{pmatrix}
\alpha_k ( g_k\beta_k + g_q^* \beta_q^*) & \alpha_k (  g_k \beta_k - g_q^* \beta_q^*) \\
\alpha_k ( -g_k \beta_k + g_q^* \beta_q^*) & \alpha_k (-g_k\beta_k - g_q^* \beta_q^*)
\end{pmatrix}
\begin{pmatrix}
\nu_q \\
\chi_q
\end{pmatrix}. \\
\hat{g}_{2n+1} &= \frac{ 1}{N/2} \sum_{kq\sigma} x_{2n+1} e^{-i(k-q)(2n+1)a} \begin{pmatrix}
\hc{\nu}_k & \hc{\chi}_k
\end{pmatrix}
\begin{pmatrix}
\alpha_k ( h_q^* \beta_k + h_k \beta_q^*) & \alpha_k (  h_q^* \beta_k - h_k \beta_q^*) \\
\alpha_k ( - h_q^* \beta_k + h_k \beta_q^*) & \alpha_k (- h_q^* \beta_k - h_k \beta_q^*)
\end{pmatrix}
\begin{pmatrix}
\nu_q \\
\chi_q
\end{pmatrix}.
\end{align}
\end{widetext}

As in the linear SSH model case, we only need a single entry from these matrices at half-filling. Moreover, defining $\delta K_{nm}^{(2)}$ as the second-order corrections to the $n,m$-binding atomic spring, by analogy we also have

\begin{widetext}
\begin{equation}\label{eq:delta_knm_2}
\delta K_{nm}^{(2)} = \frac{\partial^2 F_e^{(2)}}{\partial x_n \partial x_m} 
=  \bra{\Psi_0} \hat{g}_n \frac{(1 - \ket{\Psi_0} \bra{\Psi_0})}{E_0 - \hat{H}_{\text{unper}}}  \hat{g}_m \ket{\Psi_0} + \bra{\Psi_0} \hat{g}_m \frac{(1 - \ket{\Psi_0} \bra{\Psi_0})}{E_0 - \hat{H}_{\text{unper}}}  \hat{g}_n \ket{\Psi_0}.
\end{equation}
\end{widetext}
It is here that we have to work our way through several cases depending on the even/odd character of $n,m$. First, catalog the halves of the perturbative expressions for the even-odd cases.

\textbf{Even $m$}: 

\begin{multline}
\frac{(1 - \ket{\Psi_0} \bra{\Psi_0})}{E_0 - \hat{H}_{\text{unper}}}  \frac{\partial \hat{V}}{\partial x_{2m}} \ket{\Psi_0} = \\
= \frac{1}{N/2} \sum_{kq\sigma} \frac{e^{-i(k-q)(2m)a}}{E_k + E_q} \alpha_k(g_k\beta_k - g_q^* \beta_q^*) \hc{\chi}_k \nu_q \ket{\Psi_0}.
\end{multline}

\textbf{Odd $m$}: 

\begin{multline}
\frac{(1 - \ket{\Psi_0} \bra{\Psi_0})}{E_0 - \hat{H}_{\text{unper}}}  \frac{\partial \hat{V}}{\partial x_{2m+1}} \ket{\Psi_0} = \\
= \frac{1}{N/2} \sum_{kq\sigma} \frac{e^{-i(k-q)(2m+1)a}}{E_k + E_q} \alpha_k(h_q^*\beta_k - h_k \beta_q^*) \hc{\chi}_k \nu_q \ket{\Psi_0}.
\end{multline}

Now for corresponding left-hand sides. The main difference is that now we keep the $\hc{\nu} \chi$ combination. We immediately have

\textbf{Even $m$}: 

\begin{multline}
\bra{\Psi_0} \frac{\partial \hat{V}}{\partial x_{2m}} = \frac{1}{N/2} \sum_{k'q'\sigma}  e^{-i(k'-q')(2m)a} \times \\
\times \alpha_{k'}(g_{k'}\beta_{k'} - g_{q'}^* \beta_{q'}^*) \bra{\Psi_0} \hc{\nu}_{k'} \chi_{q'}.
\end{multline}

When connected with the appropriate right-hand side, the equality $q' = k, k' = q$ is ensured. Hence

\begin{multline}
\bra{\Psi_0} \frac{\partial \hat{V}}{\partial x_{2m}} = \frac{1}{N/2} \sum_{kq\sigma}  e^{+i(k-q)(2m)a} \times \\
\times \alpha_{q}(g_{q}\beta_{q} - g_{k}^* \beta_{k}^*) \bra{\Psi_0} \hc{\nu}_{q} \chi_{k}.
\end{multline}

The only thing that changes for odd $m$ is the prefactor:

\textbf{Odd $m$}: 

\begin{multline}
\bra{\Psi_0} \frac{\partial \hat{V}}{\partial x_{2m+1}} = \frac{1}{N/2} \sum_{kq\sigma}  e^{+i(k-q)(2m+1)a} \times \\
\times \alpha_{q}(h_{k}^*\beta_{q} - h_{q} \beta_{k}^*) \bra{\Psi_0} \hc{\nu}_{q} \chi_{k}.
\end{multline}

Now assemble the full expressions:

\textbf{Case 1: $n$ even, $m$ even}.

\begin{multline}
\delta K_{2n,2m} = \left(\frac{1}{N/2} \right)^2  \sum_{kq\sigma} \left( \frac{e^{-i(k-q)(n-m)a}}{E_k + E_q} \right) (\alpha_k^2) \times \\
\times \left\{ (-g_k \beta_k + g_q^* \beta_q^*) (-g_k^* \beta_k^* + g_q \beta_q) \right\} + (n\rightleftarrows m) = \\
= 2\left(\frac{2}{N} \right)^2  \sum_{kq} \left( \frac{\cos[ (k-q)(n-m)a ]}{E_k + E_q} \right) \times \\
\times \left|-g_k \beta_k + g_q^* \beta_q^*\right|^2
\end{multline}  

\textbf{Case 2: $n$ odd, $m$ odd}.

\begin{multline}
\delta K_{2n+1,2m+1} =  2\left(\frac{2}{N} \right)^2 \times \\
\times \sum_{kq} \left( \frac{\cos[ (k-q)(2n-2m)a ]}{E_k + E_q} \right) \left| -h_q^* \beta_k + h_k \beta_q^* \right|^2.
\end{multline}

\textbf{Case 3: $n$ even, $m$ odd}.

\begin{multline}
\delta K_{2n,2m+1} =  2\left(\frac{1}{N/2} \right)^2  \sum_{kq} \left( \frac{ e^{-i(k-q)(2m-2n+1)a}}{E_k + E_q} \right) \times \\
\times \left\{ (g_q\beta_q - g_k^* \beta_k^*)(h_q^* \beta_k - h_k \beta_q^*) \right\} + \text{h.c}.
\end{multline}

\textbf{Case 4: $n$ odd, $m$ even}.

\begin{multline}
\delta K_{nm} =  \left(\frac{1}{N/2} \right)^2  \sum_{kq} \left( \frac{ e^{-i(k-q)(2m-2n-1)a}}{E_k + E_q} \right) \times \\ \times \left\{ (h_k^* \beta_q - h_q \beta_k^*)(g_k\beta_k - g_q^* \beta_q^*) \right\} + \text{h.c}.
\end{multline}

The subsequent calculation of the phonon spectrum is carried out exactly as for the linear case. 

\section{Diagonalizing the ionic Hamiltonian}
\label{app:ionic-diag}

Starting from the Hamiltonian in Eq. \ref{eq:ham_ion_final}, we use the fact that the dispersions of quantum Hamiltonians and equivalent classical Hamiltonians are identical, and write down the corresponding classical Lagrangian

\begin{widetext}
\begin{multline}\label{eq:lag_ion_final}
\mathcal{L}_{i,\text{har}}^{\text{(BO)}} = \frac{M}{2} \sum_n  (\dot{x}_{n}^2 + \dot{X}_n^2) - \frac{1}{2} \sum_n \Big( Z_0 (x_n^2 + X_n^2) + Y_+ x_n X_n + \\
+ Y_- X_n x_{n+1} + Z_2 (x_n x_{n+1} + X_n X_{n+1} ) + \\
+ \sum_{\delta \geq 1} ( Z_{2+2\delta} x_n x_{n+1+\delta} + Z_{1+2\delta} x_n X_{n+\delta}
+ Z_{2+2\delta} X_n X_{n+1+\delta} + Z_{1+2\delta} X_n x_{n+\delta}  ) \Big) + E_{e}(\{R_i^0\}).
\end{multline}

The Euler-Lagrange equations have the form

\begin{multline}
M \ddot{x}_n = - Z_0 x_n - \Big(Y_+ X_n + Y_- X_{n-1} + \sum_{\delta \geq 1} Z_{1+2\delta} [X_{n+\delta} + X_{n-1-\delta}] \Big) - \\
- Z(x_{n+1} + x_{n-1}) - \sum_{\delta \geq 1} Z_{2+2\delta} (x_{n+1+\delta} + x_{n-1-\delta}),
\end{multline}
\begin{multline}
M \ddot{X}_n = - Z_0 X_n - \Big(Y_+ x_n + Y_- x_{n+1}  + \sum_{\delta \geq 1} Z_{1+2\delta} [x_{n+1+\delta} + x_{n-\delta}] \Big) -\\
- Z(X_{n+1} + X_{n-1}) - \sum_{\delta \geq 1} Z_{2+2\delta} (X_{n+1+\delta} + X_{n-1-\delta}).
\end{multline}

Using the standard Fourier transform
\begin{equation}
x_n = \frac{1}{\sqrt{N/2}} \sum_q e^{-2iqna-i\omega t} x_q, \quad X_n = \frac{1}{\sqrt{N/2}} \sum_q e^{-2iqna-i\omega t} X_q,
\end{equation}
 we find the equations (define $\cos[(x)qa] \equiv c_{x}$ for brevity)

\begin{equation}
\begin{pmatrix}
\frac{1}{M} \Big(Z_0 + 2 Z c_2 + 2\sum_{\delta \geq 1} Z_{2+2\delta} c_{2+2\delta} \Big) - \omega^2 & \frac{1}{M}(Y_+ + Y_- e^{2iqa} + 2e^{iqa}\sum_{\delta \geq 1} Z_{1+2\delta}  c_{1+2\delta} ) \\
\frac{1}{M} (Y_+ + Y_- e^{-2iqa} + 2e^{-iqa}\sum_{\delta \geq 1} Z_{1+2\delta}  c_{1+2\delta}) & \frac{1}{M} \Big( Z_0 + 2 Z c_2 + 2 \sum_{\delta \geq 1} Z_{2+2\delta} c_{2+2\delta}  \Big) - \omega^2
\end{pmatrix} 
\begin{pmatrix}
x_q \\
X_q
\end{pmatrix} = 0.
\end{equation}
\end{widetext}

The dispersion reported in the text is obtained from diagonalizing this matrix.

\bibliographystyle{apsrev4-2}

\begin{thebibliography}{38}%
	\makeatletter
	\providecommand \@ifxundefined [1]{%
		\@ifx{#1\undefined}
	}%
	\providecommand \@ifnum [1]{%
		\ifnum #1\expandafter \@firstoftwo
		\else \expandafter \@secondoftwo
		\fi
	}%
	\providecommand \@ifx [1]{%
		\ifx #1\expandafter \@firstoftwo
		\else \expandafter \@secondoftwo
		\fi
	}%
	\providecommand \natexlab [1]{#1}%
	\providecommand \enquote  [1]{``#1''}%
	\providecommand \bibnamefont  [1]{#1}%
	\providecommand \bibfnamefont [1]{#1}%
	\providecommand \citenamefont [1]{#1}%
	\providecommand \href@noop [0]{\@secondoftwo}%
	\providecommand \href [0]{\begingroup \@sanitize@url \@href}%
	\providecommand \@href[1]{\@@startlink{#1}\@@href}%
	\providecommand \@@href[1]{\endgroup#1\@@endlink}%
	\providecommand \@sanitize@url [0]{\catcode `\\12\catcode `\$12\catcode
		`\&12\catcode `\#12\catcode `\^12\catcode `\_12\catcode `\%12\relax}%
	\providecommand \@@startlink[1]{}%
	\providecommand \@@endlink[0]{}%
	\providecommand \url  [0]{\begingroup\@sanitize@url \@url }%
	\providecommand \@url [1]{\endgroup\@href {#1}{\urlprefix }}%
	\providecommand \urlprefix  [0]{URL }%
	\providecommand \Eprint [0]{\href }%
	\providecommand \doibase [0]{https://doi.org/}%
	\providecommand \selectlanguage [0]{\@gobble}%
	\providecommand \bibinfo  [0]{\@secondoftwo}%
	\providecommand \bibfield  [0]{\@secondoftwo}%
	\providecommand \translation [1]{[#1]}%
	\providecommand \BibitemOpen [0]{}%
	\providecommand \bibitemStop [0]{}%
	\providecommand \bibitemNoStop [0]{.\EOS\space}%
	\providecommand \EOS [0]{\spacefactor3000\relax}%
	\providecommand \BibitemShut  [1]{\csname bibitem#1\endcsname}%
	\let\auto@bib@innerbib\@empty
	\bibitem [{\citenamefont {Lennard-Jones}(1937)}]{Lennard-Jones1937}%
	\BibitemOpen
	\bibfield  {author} {\bibinfo {author} {\bibfnamefont {J.~E.}\ \bibnamefont
			{Lennard-Jones}},\ }\href {https://doi.org/10.1098/rspa.1937.0020} {\bibfield
		{journal} {\bibinfo  {journal} {P. Roy. Soc. A-Math. Phy.}\ }\textbf
		{\bibinfo {volume} {158}},\ \bibinfo {pages} {280} (\bibinfo {year}
		{1937})}\BibitemShut {NoStop}%
	\bibitem [{\citenamefont {Coulson}(1939)}]{Coulson1939}%
	\BibitemOpen
	\bibfield  {author} {\bibinfo {author} {\bibfnamefont {C.~A.}\ \bibnamefont
			{Coulson}},\ }\href {https://doi.org/10.1098/rspa.1939.0006} {\bibfield
		{journal} {\bibinfo  {journal} {P. Roy. Soc. A-Math. Phy.}\ }\textbf
		{\bibinfo {volume} {169}},\ \bibinfo {pages} {413} (\bibinfo {year}
		{1939})}\BibitemShut {NoStop}%
	\bibitem [{\citenamefont {Kuhn}(1937)}]{Kuhn1937}%
	\BibitemOpen
	\bibfield  {author} {\bibinfo {author} {\bibfnamefont {R.}~\bibnamefont
			{Kuhn}},\ }\href {https://doi.org/10.1002/ange.19370503402} {\bibfield
		{journal} {\bibinfo  {journal} {Angew. Chem. Int. Edit.}\ }\textbf {\bibinfo
			{volume} {50}},\ \bibinfo {pages} {703} (\bibinfo {year} {1937})}\BibitemShut
	{NoStop}%
	\bibitem [{\citenamefont {Peierls}(2001)}]{Peierls2001}%
	\BibitemOpen
	\bibfield  {author} {\bibinfo {author} {\bibfnamefont {R.~E.}\ \bibnamefont
			{Peierls}},\ }\href@noop {} {\emph {\bibinfo {title} {Quantum Theory of
				Solids}}}\ (\bibinfo  {publisher} {Claredon Press},\ \bibinfo {address}
	{Oxford},\ \bibinfo {year} {2001})\BibitemShut {NoStop}%
	\bibitem [{\citenamefont {Ooshika}(1957)}]{Ooshika1957}%
	\BibitemOpen
	\bibfield  {author} {\bibinfo {author} {\bibfnamefont {Y.}~\bibnamefont
			{Ooshika}},\ }\href {https://doi.org/10.1143/JPSJ.12.1246} {\bibfield
		{journal} {\bibinfo  {journal} {J. Phys. Soc. Jpn.}\ }\textbf {\bibinfo
			{volume} {12}},\ \bibinfo {pages} {1246} (\bibinfo {year}
		{1957})}\BibitemShut {NoStop}%
	\bibitem [{\citenamefont {Longuet-Higgins}\ and\ \citenamefont
		{Salem}(1959)}]{Longuet-Higgins1959}%
	\BibitemOpen
	\bibfield  {author} {\bibinfo {author} {\bibfnamefont {H.~C.}\ \bibnamefont
			{Longuet-Higgins}}\ and\ \bibinfo {author} {\bibfnamefont {L.}~\bibnamefont
			{Salem}},\ }\href {https://doi.org/10.1098/rspa.1959.0100} {\bibfield
		{journal} {\bibinfo  {journal} {P. Roy. Soc. A-Math. Phy.}\ }\textbf
		{\bibinfo {volume} {251}},\ \bibinfo {pages} {172} (\bibinfo {year}
		{1959})}\BibitemShut {NoStop}%
	\bibitem [{\citenamefont {Su}\ \emph {et~al.}(1979)\citenamefont {Su},
		\citenamefont {Schrieffer},\ and\ \citenamefont {Heeger}}]{Su1979}%
	\BibitemOpen
	\bibfield  {author} {\bibinfo {author} {\bibfnamefont {W.~P.}\ \bibnamefont
			{Su}}, \bibinfo {author} {\bibfnamefont {J.~R.}\ \bibnamefont {Schrieffer}},\
		and\ \bibinfo {author} {\bibfnamefont {A.~J.}\ \bibnamefont {Heeger}},\
	}\href {https://doi.org/10.1103/PhysRevLett.42.1698} {\bibfield  {journal}
		{\bibinfo  {journal} {Phys. Rev. Lett.}\ }\textbf {\bibinfo {volume} {42}},\
		\bibinfo {pages} {1698} (\bibinfo {year} {1979})}\BibitemShut {NoStop}%
	\bibitem [{\citenamefont {Su}(1982)}]{Su1982}%
	\BibitemOpen
	\bibfield  {author} {\bibinfo {author} {\bibfnamefont {W.}~\bibnamefont
			{Su}},\ }\href {https://doi.org/10.1016/0038-1098(82)90628-7} {\bibfield
		{journal} {\bibinfo  {journal} {Solid State Commun.}\ }\textbf {\bibinfo
			{volume} {42}},\ \bibinfo {pages} {497} (\bibinfo {year} {1982})}\BibitemShut
	{NoStop}%
	\bibitem [{\citenamefont {Nakahara}\ and\ \citenamefont
		{Maki}(1982)}]{Nakahara1982}%
	\BibitemOpen
	\bibfield  {author} {\bibinfo {author} {\bibfnamefont {M.}~\bibnamefont
			{Nakahara}}\ and\ \bibinfo {author} {\bibfnamefont {K.}~\bibnamefont
			{Maki}},\ }\href {https://doi.org/10.1103/PhysRevB.25.7789} {\bibfield
		{journal} {\bibinfo  {journal} {Phys. Rev. B}\ }\textbf {\bibinfo {volume}
			{25}},\ \bibinfo {pages} {7789} (\bibinfo {year} {1982})}\BibitemShut
	{NoStop}%
	\bibitem [{\citenamefont {Fradkin}\ and\ \citenamefont
		{Hirsch}(1983)}]{Fradkin1983}%
	\BibitemOpen
	\bibfield  {author} {\bibinfo {author} {\bibfnamefont {E.}~\bibnamefont
			{Fradkin}}\ and\ \bibinfo {author} {\bibfnamefont {J.~E.}\ \bibnamefont
			{Hirsch}},\ }\href {https://doi.org/10.1103/PhysRevB.27.1680} {\bibfield
		{journal} {\bibinfo  {journal} {Phys. Rev. B}\ }\textbf {\bibinfo {volume}
			{27}},\ \bibinfo {pages} {1680} (\bibinfo {year} {1983})}\BibitemShut
	{NoStop}%
	\bibitem [{\citenamefont {Sengupta}\ \emph {et~al.}(2003)\citenamefont
		{Sengupta}, \citenamefont {Sandvik},\ and\ \citenamefont
		{Campbell}}]{Sengupta2003}%
	\BibitemOpen
	\bibfield  {author} {\bibinfo {author} {\bibfnamefont {P.}~\bibnamefont
			{Sengupta}}, \bibinfo {author} {\bibfnamefont {A.~W.}\ \bibnamefont
			{Sandvik}},\ and\ \bibinfo {author} {\bibfnamefont {D.~K.}\ \bibnamefont
			{Campbell}},\ }\href {https://doi.org/10.1103/PhysRevB.67.245103} {\bibfield
		{journal} {\bibinfo  {journal} {Phys. Rev. B}\ }\textbf {\bibinfo {volume}
			{67}},\ \bibinfo {pages} {245103} (\bibinfo {year} {2003})}\BibitemShut
	{NoStop}%
	\bibitem [{\citenamefont {Barford}\ and\ \citenamefont
		{Bursill}(2006)}]{Barford2006}%
	\BibitemOpen
	\bibfield  {author} {\bibinfo {author} {\bibfnamefont {W.}~\bibnamefont
			{Barford}}\ and\ \bibinfo {author} {\bibfnamefont {R.~J.}\ \bibnamefont
			{Bursill}},\ }\href {https://doi.org/10.1103/PhysRevB.73.045106} {\bibfield
		{journal} {\bibinfo  {journal} {Phys. Rev. B}\ }\textbf {\bibinfo {volume}
			{73}},\ \bibinfo {pages} {045106} (\bibinfo {year} {2006})}\BibitemShut
	{NoStop}%
	\bibitem [{\citenamefont {Pearson}\ \emph {et~al.}(2011)\citenamefont
		{Pearson}, \citenamefont {Barford},\ and\ \citenamefont
		{Bursill}}]{Pearson2011}%
	\BibitemOpen
	\bibfield  {author} {\bibinfo {author} {\bibfnamefont {C.~J.}\ \bibnamefont
			{Pearson}}, \bibinfo {author} {\bibfnamefont {W.}~\bibnamefont {Barford}},\
		and\ \bibinfo {author} {\bibfnamefont {R.~J.}\ \bibnamefont {Bursill}},\
	}\href {https://doi.org/10.1103/PhysRevB.83.195105} {\bibfield  {journal}
		{\bibinfo  {journal} {Phys. Rev. B}\ }\textbf {\bibinfo {volume} {83}},\
		\bibinfo {pages} {195105} (\bibinfo {year} {2011})}\BibitemShut {NoStop}%
	\bibitem [{\citenamefont {Weber}\ \emph {et~al.}(2015)\citenamefont {Weber},
		\citenamefont {Assaad},\ and\ \citenamefont {Hohenadler}}]{Weber2015}%
	\BibitemOpen
	\bibfield  {author} {\bibinfo {author} {\bibfnamefont {M.}~\bibnamefont
			{Weber}}, \bibinfo {author} {\bibfnamefont {F.~F.}\ \bibnamefont {Assaad}},\
		and\ \bibinfo {author} {\bibfnamefont {M.}~\bibnamefont {Hohenadler}},\
	}\href {https://doi.org/10.1103/PhysRevB.91.245147} {\bibfield  {journal}
		{\bibinfo  {journal} {Phys. Rev. B}\ }\textbf {\bibinfo {volume} {91}},\
		\bibinfo {pages} {245147} (\bibinfo {year} {2015})}\BibitemShut {NoStop}%
	\bibitem [{\citenamefont {Barford}\ \emph {et~al.}(2002)\citenamefont
		{Barford}, \citenamefont {Bursill},\ and\ \citenamefont
		{Lavrentiev}}]{Barford2002}%
	\BibitemOpen
	\bibfield  {author} {\bibinfo {author} {\bibfnamefont {W.}~\bibnamefont
			{Barford}}, \bibinfo {author} {\bibfnamefont {R.~J.}\ \bibnamefont
			{Bursill}},\ and\ \bibinfo {author} {\bibfnamefont {M.~Y.}\ \bibnamefont
			{Lavrentiev}},\ }\href {https://doi.org/10.1103/PhysRevB.65.075107}
	{\bibfield  {journal} {\bibinfo  {journal} {Phys. Rev. B}\ }\textbf {\bibinfo
			{volume} {65}},\ \bibinfo {pages} {075107} (\bibinfo {year}
		{2002})}\BibitemShut {NoStop}%
	\bibitem [{\citenamefont {Shi}\ \emph {et~al.}(2016)\citenamefont {Shi},
		\citenamefont {Rohringer}, \citenamefont {Suenaga}, \citenamefont {Niimi},
		\citenamefont {Kotakoski}, \citenamefont {Meyer}, \citenamefont {Peterlik},
		\citenamefont {Wanko}, \citenamefont {Cahangirov}, \citenamefont {Rubio},
		\citenamefont {Lapin}, \citenamefont {Novotny}, \citenamefont {Ayala},\ and\
		\citenamefont {Pichler}}]{Shi2016}%
	\BibitemOpen
	\bibfield  {author} {\bibinfo {author} {\bibfnamefont {L.}~\bibnamefont
			{Shi}}, \bibinfo {author} {\bibfnamefont {P.}~\bibnamefont {Rohringer}},
		\bibinfo {author} {\bibfnamefont {K.}~\bibnamefont {Suenaga}}, \bibinfo
		{author} {\bibfnamefont {Y.}~\bibnamefont {Niimi}}, \bibinfo {author}
		{\bibfnamefont {J.}~\bibnamefont {Kotakoski}}, \bibinfo {author}
		{\bibfnamefont {J.~C.}\ \bibnamefont {Meyer}}, \bibinfo {author}
		{\bibfnamefont {H.}~\bibnamefont {Peterlik}}, \bibinfo {author}
		{\bibfnamefont {M.}~\bibnamefont {Wanko}}, \bibinfo {author} {\bibfnamefont
			{S.}~\bibnamefont {Cahangirov}}, \bibinfo {author} {\bibfnamefont
			{A.}~\bibnamefont {Rubio}}, \bibinfo {author} {\bibfnamefont {Z.~J.}\
			\bibnamefont {Lapin}}, \bibinfo {author} {\bibfnamefont {L.}~\bibnamefont
			{Novotny}}, \bibinfo {author} {\bibfnamefont {P.}~\bibnamefont {Ayala}},\
		and\ \bibinfo {author} {\bibfnamefont {T.}~\bibnamefont {Pichler}},\ }\href
	{https://doi.org/10.1038/nmat4617} {\bibfield  {journal} {\bibinfo  {journal}
			{Nat. Mater.}\ }\textbf {\bibinfo {volume} {15}},\ \bibinfo {pages} {634}
		(\bibinfo {year} {2016})}\BibitemShut {NoStop}%
	\bibitem [{\citenamefont {Hudson}\ and\ \citenamefont
		{Allis}(2013)}]{Hudson2013}%
	\BibitemOpen
	\bibfield  {author} {\bibinfo {author} {\bibfnamefont {B.~S.}\ \bibnamefont
			{Hudson}}\ and\ \bibinfo {author} {\bibfnamefont {D.~G.}\ \bibnamefont
			{Allis}},\ }\href {https://doi.org/10.1016/j.molstruc.2012.07.051} {\bibfield
		{journal} {\bibinfo  {journal} {J. Mol. Struct.}\ }\textbf {\bibinfo
			{volume} {1032}},\ \bibinfo {pages} {78} (\bibinfo {year}
		{2013})}\BibitemShut {NoStop}%
	\bibitem [{\citenamefont {Artyukhov}\ \emph {et~al.}(2014)\citenamefont
		{Artyukhov}, \citenamefont {Liu},\ and\ \citenamefont
		{Yakobson}}]{Artyukhov2014}%
	\BibitemOpen
	\bibfield  {author} {\bibinfo {author} {\bibfnamefont {V.~I.}\ \bibnamefont
			{Artyukhov}}, \bibinfo {author} {\bibfnamefont {M.}~\bibnamefont {Liu}},\
		and\ \bibinfo {author} {\bibfnamefont {B.~I.}\ \bibnamefont {Yakobson}},\
	}\href {https://doi.org/10.1021/nl5017317} {\bibfield  {journal} {\bibinfo
			{journal} {Nano Lett.}\ }\textbf {\bibinfo {volume} {14}},\ \bibinfo {pages}
		{4224} (\bibinfo {year} {2014})}\BibitemShut {NoStop}%
	\bibitem [{\citenamefont {{La Torre}}\ \emph {et~al.}(2015)\citenamefont {{La
				Torre}}, \citenamefont {Botello-Mendez}, \citenamefont {Baaziz},
		\citenamefont {Charlier},\ and\ \citenamefont {Banhart}}]{LaTorre2015}%
	\BibitemOpen
	\bibfield  {author} {\bibinfo {author} {\bibfnamefont {A.}~\bibnamefont {{La
					Torre}}}, \bibinfo {author} {\bibfnamefont {A.}~\bibnamefont
			{Botello-Mendez}}, \bibinfo {author} {\bibfnamefont {W.}~\bibnamefont
			{Baaziz}}, \bibinfo {author} {\bibfnamefont {J.~C.}\ \bibnamefont
			{Charlier}},\ and\ \bibinfo {author} {\bibfnamefont {F.}~\bibnamefont
			{Banhart}},\ }\href {https://doi.org/10.1038/ncomms7636} {\bibfield
		{journal} {\bibinfo  {journal} {Nat. Commun.}\ }\textbf {\bibinfo {volume}
			{6}},\ \bibinfo {pages} {6636} (\bibinfo {year} {2015})}\BibitemShut
	{NoStop}%
	\bibitem [{\citenamefont {{Ben Romdhane}}\ \emph {et~al.}(2017)\citenamefont
		{{Ben Romdhane}}, \citenamefont {Adjizian}, \citenamefont {Charlier},\ and\
		\citenamefont {Banhart}}]{BenRomdhane2017}%
	\BibitemOpen
	\bibfield  {author} {\bibinfo {author} {\bibfnamefont {F.}~\bibnamefont {{Ben
					Romdhane}}}, \bibinfo {author} {\bibfnamefont {J.-J.}\ \bibnamefont
			{Adjizian}}, \bibinfo {author} {\bibfnamefont {J.-C.}\ \bibnamefont
			{Charlier}},\ and\ \bibinfo {author} {\bibfnamefont {F.}~\bibnamefont
			{Banhart}},\ }\href {https://doi.org/10.1016/j.carbon.2017.06.039} {\bibfield
		{journal} {\bibinfo  {journal} {Carbon}\ }\textbf {\bibinfo {volume}
			{122}},\ \bibinfo {pages} {92} (\bibinfo {year} {2017})}\BibitemShut
	{NoStop}%
	\bibitem [{\citenamefont {Hudson}(2018)}]{Hudson2018}%
	\BibitemOpen
	\bibfield  {author} {\bibinfo {author} {\bibfnamefont {B.}~\bibnamefont
			{Hudson}},\ }\href {https://doi.org/10.3390/ma11020242} {\bibfield  {journal}
		{\bibinfo  {journal} {Materials}\ }\textbf {\bibinfo {volume} {11}},\
		\bibinfo {pages} {242} (\bibinfo {year} {2018})}\BibitemShut {NoStop}%
	\bibitem [{\citenamefont {Ovchinnikov}\ \emph {et~al.}(1982)\citenamefont
		{Ovchinnikov}, \citenamefont {Belinskii}, \citenamefont {Misurkin},\ and\
		\citenamefont {Ukrainskii}}]{Ovchinnikov1982}%
	\BibitemOpen
	\bibfield  {author} {\bibinfo {author} {\bibfnamefont {A.~A.}\ \bibnamefont
			{Ovchinnikov}}, \bibinfo {author} {\bibfnamefont {A.~E.}\ \bibnamefont
			{Belinskii}}, \bibinfo {author} {\bibfnamefont {I.~A.}\ \bibnamefont
			{Misurkin}},\ and\ \bibinfo {author} {\bibfnamefont {I.~I.}\ \bibnamefont
			{Ukrainskii}},\ }\href {https://doi.org/10.1002/qua.560220409} {\bibfield
		{journal} {\bibinfo  {journal} {Int. J. Quantum Chem.}\ }\textbf {\bibinfo
			{volume} {22}},\ \bibinfo {pages} {761} (\bibinfo {year} {1982})}\BibitemShut
	{NoStop}%
	\bibitem [{\citenamefont {Zhu}\ \emph {et~al.}(1992)\citenamefont {Zhu},
		\citenamefont {Fischer}, \citenamefont {Zusok},\ and\ \citenamefont
		{Roth}}]{Zhu1992}%
	\BibitemOpen
	\bibfield  {author} {\bibinfo {author} {\bibfnamefont {Q.}~\bibnamefont
			{Zhu}}, \bibinfo {author} {\bibfnamefont {J.~E.}\ \bibnamefont {Fischer}},
		\bibinfo {author} {\bibfnamefont {R.}~\bibnamefont {Zusok}},\ and\ \bibinfo
		{author} {\bibfnamefont {S.}~\bibnamefont {Roth}},\ }\href
	{https://doi.org/10.1016/0038-1098(92)90832-T} {\bibfield  {journal}
		{\bibinfo  {journal} {Solid State Commun.}\ }\textbf {\bibinfo {volume}
			{83}},\ \bibinfo {pages} {179} (\bibinfo {year} {1992})}\BibitemShut
	{NoStop}%
	\bibitem [{\citenamefont {Yannoni}\ and\ \citenamefont
		{Clarke}(1983)}]{Yannoni1983}%
	\BibitemOpen
	\bibfield  {author} {\bibinfo {author} {\bibfnamefont {C.~S.}\ \bibnamefont
			{Yannoni}}\ and\ \bibinfo {author} {\bibfnamefont {T.~C.}\ \bibnamefont
			{Clarke}},\ }\href {https://doi.org/10.1103/PhysRevLett.51.1191} {\bibfield
		{journal} {\bibinfo  {journal} {Phys. Rev. Lett.}\ }\textbf {\bibinfo
			{volume} {51}},\ \bibinfo {pages} {1191} (\bibinfo {year}
		{1983})}\BibitemShut {NoStop}%
	\bibitem [{\citenamefont {Schen}\ \emph {et~al.}(1988)\citenamefont {Schen},
		\citenamefont {Chien}, \citenamefont {Perrin}, \citenamefont {Lefrant},\ and\
		\citenamefont {Mulazzi}}]{Schen1988}%
	\BibitemOpen
	\bibfield  {author} {\bibinfo {author} {\bibfnamefont {M.~A.}\ \bibnamefont
			{Schen}}, \bibinfo {author} {\bibfnamefont {J.~C.~W.}\ \bibnamefont {Chien}},
		\bibinfo {author} {\bibfnamefont {E.}~\bibnamefont {Perrin}}, \bibinfo
		{author} {\bibfnamefont {S.}~\bibnamefont {Lefrant}},\ and\ \bibinfo {author}
		{\bibfnamefont {E.}~\bibnamefont {Mulazzi}},\ }\href
	{https://doi.org/10.1063/1.455248} {\bibfield  {journal} {\bibinfo  {journal}
			{J. Chem. Phys.}\ }\textbf {\bibinfo {volume} {89}},\ \bibinfo {pages} {7615}
		(\bibinfo {year} {1988})}\BibitemShut {NoStop}%
	\bibitem [{\citenamefont {Dincă}\ \emph {et~al.}(2020)\citenamefont {Dincă},
		\citenamefont {Allis}, \citenamefont {Moskowitz}, \citenamefont {Sponsler},\
		and\ \citenamefont {Hudson}}]{Dinca2020}%
	\BibitemOpen
	\bibfield  {author} {\bibinfo {author} {\bibfnamefont {S.~A.}\ \bibnamefont
			{Dincă}}, \bibinfo {author} {\bibfnamefont {D.~G.}\ \bibnamefont {Allis}},
		\bibinfo {author} {\bibfnamefont {M.~D.}\ \bibnamefont {Moskowitz}}, \bibinfo
		{author} {\bibfnamefont {M.~B.}\ \bibnamefont {Sponsler}},\ and\ \bibinfo
		{author} {\bibfnamefont {B.~S.}\ \bibnamefont {Hudson}},\ }\href
	{https://doi.org/10.1021/acs.chemmater.9b03644} {\bibfield  {journal}
		{\bibinfo  {journal} {Chem. Mater.}\ }\textbf {\bibinfo {volume} {32}},\
		\bibinfo {pages} {1769} (\bibinfo {year} {2020})}\BibitemShut {NoStop}%
	\bibitem [{\citenamefont {Ovchinnikov}\ \emph {et~al.}(1973)\citenamefont
		{Ovchinnikov}, \citenamefont {Ukrainskiĭ},\ and\ \citenamefont
		{Kventsel'}}]{Ovchinnikov1973}%
	\BibitemOpen
	\bibfield  {author} {\bibinfo {author} {\bibfnamefont {A.~A.}\ \bibnamefont
			{Ovchinnikov}}, \bibinfo {author} {\bibfnamefont {I.~I.}\ \bibnamefont
			{Ukrainskiĭ}},\ and\ \bibinfo {author} {\bibfnamefont {G.~V.}\ \bibnamefont
			{Kventsel'}},\ }\href {https://doi.org/10.1070/PU1973v015n05ABEH005011}
	{\bibfield  {journal} {\bibinfo  {journal} {Sov. Phys. Uspekhi}\ }\textbf
		{\bibinfo {volume} {15}},\ \bibinfo {pages} {575} (\bibinfo {year}
		{1973})}\BibitemShut {NoStop}%
	\bibitem [{\citenamefont {Schulz}(1978)}]{Schulz1978}%
	\BibitemOpen
	\bibfield  {author} {\bibinfo {author} {\bibfnamefont {H.~J.}\ \bibnamefont
			{Schulz}},\ }\href {https://doi.org/10.1103/PhysRevB.18.5756} {\bibfield
		{journal} {\bibinfo  {journal} {Phys. Rev. B}\ }\textbf {\bibinfo {volume}
			{18}},\ \bibinfo {pages} {5756} (\bibinfo {year} {1978})}\BibitemShut
	{NoStop}%
	\bibitem [{\citenamefont {Souvatzis}\ \emph {et~al.}(2008)\citenamefont
		{Souvatzis}, \citenamefont {Eriksson}, \citenamefont {Katsnelson},\ and\
		\citenamefont {Rudin}}]{Souvatzis2008}%
	\BibitemOpen
	\bibfield  {author} {\bibinfo {author} {\bibfnamefont {P.}~\bibnamefont
			{Souvatzis}}, \bibinfo {author} {\bibfnamefont {O.}~\bibnamefont {Eriksson}},
		\bibinfo {author} {\bibfnamefont {M.}~\bibnamefont {Katsnelson}},\ and\
		\bibinfo {author} {\bibfnamefont {S.}~\bibnamefont {Rudin}},\ }\href
	{https://journals.aps.org/prl/abstract/10.1103/PhysRevLett.100.095901}
	{\bibfield  {journal} {\bibinfo  {journal} {Phys. Rev. Lett.}\ }\textbf
		{\bibinfo {volume} {100}},\ \bibinfo {pages} {095901} (\bibinfo {year}
		{2008})}\BibitemShut {NoStop}%
	\bibitem [{\citenamefont {Monacelli}\ \emph {et~al.}(2021)\citenamefont
		{Monacelli}, \citenamefont {Bianco}, \citenamefont {Cherubini}, \citenamefont
		{Calandra}, \citenamefont {Errea},\ and\ \citenamefont
		{Mauri}}]{Monacelli2021}%
	\BibitemOpen
	\bibfield  {author} {\bibinfo {author} {\bibfnamefont {L.}~\bibnamefont
			{Monacelli}}, \bibinfo {author} {\bibfnamefont {R.}~\bibnamefont {Bianco}},
		\bibinfo {author} {\bibfnamefont {M.}~\bibnamefont {Cherubini}}, \bibinfo
		{author} {\bibfnamefont {M.}~\bibnamefont {Calandra}}, \bibinfo {author}
		{\bibfnamefont {I.}~\bibnamefont {Errea}},\ and\ \bibinfo {author}
		{\bibfnamefont {F.}~\bibnamefont {Mauri}},\ }\href
	{https://iopscience.iop.org/article/10.1088/1361-648X/ac066b} {\bibfield
		{journal} {\bibinfo  {journal} {J. Phys.-Condens. Mat.}\ }\textbf {\bibinfo
			{volume} {33}},\ \bibinfo {pages} {363001} (\bibinfo {year}
		{2021})}\BibitemShut {NoStop}%
	\bibitem [{\citenamefont {Swager}(2017)}]{Swager2017}%
	\BibitemOpen
	\bibfield  {author} {\bibinfo {author} {\bibfnamefont {T.~M.}\ \bibnamefont
			{Swager}},\ }\href {https://doi.org/10.1021/acs.macromol.7b00582} {\bibfield
		{journal} {\bibinfo  {journal} {Macromolecules}\ }\textbf {\bibinfo {volume}
			{50}},\ \bibinfo {pages} {4867} (\bibinfo {year} {2017})}\BibitemShut
	{NoStop}%
	\bibitem [{\citenamefont {Born}\ and\ \citenamefont {Huang}(1954)}]{Born1954}%
	\BibitemOpen
	\bibfield  {author} {\bibinfo {author} {\bibfnamefont {M.}~\bibnamefont
			{Born}}\ and\ \bibinfo {author} {\bibfnamefont {K.}~\bibnamefont {Huang}},\
	}\href@noop {} {\emph {\bibinfo {title} {{Dynamical theory of crystal
					lattices}}}}\ (\bibinfo  {publisher} {Clarendon Press},\ \bibinfo {address}
	{Oxford},\ \bibinfo {year} {1954})\BibitemShut {NoStop}%
	\bibitem [{\citenamefont {Slater}(1951)}]{Slater1951}%
	\BibitemOpen
	\bibfield  {author} {\bibinfo {author} {\bibfnamefont {J.~C.}\ \bibnamefont
			{Slater}},\ }\href@noop {} {\emph {\bibinfo {title} {{Quantum Theory of
					Matter}}}},\ \bibinfo {edition} {1st}\ ed.\ (\bibinfo  {publisher}
	{McGraw-Hill Book Company},\ \bibinfo {address} {Cambridge, Mass.},\ \bibinfo
	{year} {1951})\BibitemShut {NoStop}%
	\bibitem [{\citenamefont {Miao}\ \emph {et~al.}(1999)\citenamefont {Miao},
		\citenamefont {Wu}, \citenamefont {Sun}, \citenamefont {Fu}, \citenamefont
		{Li},\ and\ \citenamefont {Kawazoe}}]{Miao1999}%
	\BibitemOpen
	\bibfield  {author} {\bibinfo {author} {\bibfnamefont {J.}~\bibnamefont
			{Miao}}, \bibinfo {author} {\bibfnamefont {C.}~\bibnamefont {Wu}}, \bibinfo
		{author} {\bibfnamefont {X.}~\bibnamefont {Sun}}, \bibinfo {author}
		{\bibfnamefont {R.}~\bibnamefont {Fu}}, \bibinfo {author} {\bibfnamefont
			{Z.}~\bibnamefont {Li}},\ and\ \bibinfo {author} {\bibfnamefont
			{Y.}~\bibnamefont {Kawazoe}},\ }\href
	{https://doi.org/10.1016/S0379-6779(98)01169-2} {\bibfield  {journal}
		{\bibinfo  {journal} {Synthetic Met.}\ }\textbf {\bibinfo {volume} {101}},\
		\bibinfo {pages} {314} (\bibinfo {year} {1999})}\BibitemShut {NoStop}%
	\bibitem [{\citenamefont {Fan}\ \emph {et~al.}(1988)\citenamefont {Fan},
		\citenamefont {Chang-qin},\ and\ \citenamefont {Xin}}]{Fan1988}%
	\BibitemOpen
	\bibfield  {author} {\bibinfo {author} {\bibfnamefont {L.}~\bibnamefont
			{Fan}}, \bibinfo {author} {\bibfnamefont {W.}~\bibnamefont {Chang-qin}},\
		and\ \bibinfo {author} {\bibfnamefont {S.}~\bibnamefont {Xin}},\ }\href
	{https://doi.org/10.1088/0253-6102/10/2/161} {\bibfield  {journal} {\bibinfo
			{journal} {Commun. Theor. Phys.}\ }\textbf {\bibinfo {volume} {10}},\
		\bibinfo {pages} {161} (\bibinfo {year} {1988})}\BibitemShut {NoStop}%
	\bibitem [{\citenamefont {Lynge}\ and\ \citenamefont
		{Pedersen}(2004)}]{Lynge2004}%
	\BibitemOpen
	\bibfield  {author} {\bibinfo {author} {\bibfnamefont {T.~B.}\ \bibnamefont
			{Lynge}}\ and\ \bibinfo {author} {\bibfnamefont {T.~G.}\ \bibnamefont
			{Pedersen}},\ }\href {https://doi.org/10.1002/pssb.200301987} {\bibfield
		{journal} {\bibinfo  {journal} {Physica Status Solidi B}\ }\textbf {\bibinfo
			{volume} {241}},\ \bibinfo {pages} {1005} (\bibinfo {year}
		{2004})}\BibitemShut {NoStop}%
	\bibitem [{\citenamefont {Derriche}\ and\ \citenamefont
		{Sawatzky}(2021)}]{Derriche2021}%
	\BibitemOpen
	\bibfield  {author} {\bibinfo {author} {\bibfnamefont {N.}~\bibnamefont
			{Derriche}}\ and\ \bibinfo {author} {\bibfnamefont {G.}~\bibnamefont
			{Sawatzky}},\ }\href {https://meetings.aps.org/Meeting/MAR21/Session/F20.7}
	{\bibfield  {journal} {\bibinfo  {journal} {B. Am. Phys. Soc.}\ } (\bibinfo
		{year} {2021})}\BibitemShut {NoStop}%
	\bibitem [{\citenamefont {Milani}\ \emph {et~al.}(2008)\citenamefont {Milani},
		\citenamefont {Tommasini},\ and\ \citenamefont {Zerbi}}]{Milani2008}%
	\BibitemOpen
	\bibfield  {author} {\bibinfo {author} {\bibfnamefont {A.}~\bibnamefont
			{Milani}}, \bibinfo {author} {\bibfnamefont {M.}~\bibnamefont {Tommasini}},\
		and\ \bibinfo {author} {\bibfnamefont {G.}~\bibnamefont {Zerbi}},\ }\href
	{https://doi.org/10.1063/1.2831507} {\bibfield  {journal} {\bibinfo
			{journal} {J. Chem. Phys.}\ }\textbf {\bibinfo {volume} {128}},\ \bibinfo
		{pages} {064501} (\bibinfo {year} {2008})}\BibitemShut {NoStop}%
\end{thebibliography}

\end{document}